\documentclass[prl,amsmath,amssymb,twocolumn,floatfix]{revtex4}
\usepackage{graphicx}

\begin{document}

\title{Muon spin rotation 
studies of electronic excitations and magnetism in the vortex cores
of superconductors}

\author{J. E. Sonier}

\email{jsonier@sfu.ca}

\affiliation{Department of Physics, Simon Fraser University,
Burnaby, British Columbia V5A 1S6, Canada}

\begin{abstract}
The focus of this article is on recent progress in muon spin rotation 
($\mu$SR) studies of the vortex cores in type-II superconductors.
By comparison of $\mu$SR measurements of the vortex-core size
in a variety of materials with results from techniques that directly 
probe electronic states, the effect of delocalized quasiparticles on 
the spatial variation of field in a lattice of interacting vortices
has been determined for both single-band and multi-band superconductors. 
These studies demonstrate the remarkable accuracy of 
what some still consider an exotic technique.
In recent years $\mu$SR has also been used to search for magnetism in 
and around the vortex cores of high-temperature superconductors. 
As a local probe $\mu$SR is specially suited for detecting static
or quasistatic magnetism having short-range or random spatial 
correlations. As discussed in this review, 
$\mu$SR experiments support a generic phase diagram
of competing superconducting and magnetic order parameters,
characterized by a quantum phase transition to a state where the
competing order is spatially non-uniform.          
\end{abstract}

\maketitle

\section{Introduction}
Muon spin rotation ($\mu$SR) is an experimental technique primarily 
used to measure local magnetic fields inside samples.
The discovery of high-transition temperature (high-$T_c$) superconductivity 
in 1986 brought about a rapid world-wide expansion in the use and applications
of $\mu$SR. Since then $\mu$SR has been routinely applied to 
to investigations of these and other newly discovered type-II superconductors.
The technique allows for studies in zero external magnetic field,
which combined with its sensitivity as a local probe has provided 
distinctive information on the occurrence of internal magnetism
as a co-existing or competing phase, or as a consequence of 
time-reversal symmetry breaking superconductivity. From zero-field
$\mu$SR studies of cuprates, a generic temperature-vs-doping phase
diagram has been constructed, showing the coexistence of
high-$T_c$ superconductivity with static magnetism in lightly doped
samples. Today there is still much debate on the origin of this magnetism
and its importance to the high-$T_c$ `problem'. 

The vortex state provides another avenue for investigation of type-II 
superconductors with $\mu$SR \cite{Sonier:00}. 
For many years such studies focussed 
solely on obtaining experimental information on the magnetic penetration 
depth ($\lambda$), through measurements of the muon spin depolarization 
rate ($\sigma$) resulting from the broad internal magnetic
field distribution $n(B)$ of the flux-line lattice (FLL). 
The temperature and magnetic field
dependences of $\lambda$, which in many systems can also be determined
in the Meissner phase by other techniques, reflect the pairing state 
symmetry of the superconducting carriers. 
With further advances of the $\mu$SR
method came the ability to focus attention on the properties of
the vortex cores themselves. 

The first-ever study to account for the finite size of the vortex cores in 
the analysis of $\mu$SR data was an investigation by 
Herlach {\it et al.} \cite{Herlach:90} of $n(B)$ in pure Nb single crystals.   
The measured field distributions were shown to be consistent with
numerical solutions of the microscopic BCS-Gor'kov theory.
Some years later, the magnetic field dependence of the 
vortex core size was determined from $\mu$SR measurements 
on single-crystal NbSe$_2$ \cite{Sonier:97a}. The results
confirmed earlier scanning tunnelling spectroscopy (STS) measurements
on NbSe$_2$ that showed a shrinking of the vortex cores with increasing 
magnetic field \cite{Golubov:94}. This behaviour could be attributed 
to an increased overlap of the quasiparticle states around a vortex core
with those coming from neighbouring vortices. However, these
$\mu$SR studies were more than just another means of accessing 
information obtainable by another experimental technique. 
Instead they marked the development of
a more powerful method for investigating some of the intrinsic properties
of vortex cores in type-II superconductors. 

The STS technique, which is sensitive to the electronic structure of 
the vortex cores, is limited to probing individual vortices 
near the sample surface. Near the surface the vortices spread out 
\cite{Kirtley:99, Niedermayer:99} and their properties are strongly 
influenced by surface inhomogeneities and/or defects. Today one can
study vortices immediately above or below the surface by $\mu$SR 
using low-energy (several keV) positively charged muons ($\mu^+$)
\cite{Niedermayer:99,Morenzoni:04}, or by $\beta$-detected NMR 
using low-energy radioactive ions \cite{Salman:07}.    
In contrast, the experiments of Refs.~\cite{Herlach:90,Sonier:97a} 
used energetic ($\sim \! 4$~MeV) $\mu^+$ that stop at interstitial 
or bond sites in the bulk of the sample where they directly probe the
local magnetic fields. The term ``bulk'' means that the stopping range
of these faster muons is approximately 150 mg/cm$^2$, which requires
samples $\sim \! 1$~mm thick.   
In further contrast to the STS method, 
$\mu$SR studies yield average information on the vortex cores, 
using $\sim \! 10^7$ $\mu^+$ to randomly probe the $\sim \! 10^9$
vortices in a typical size sample. 

Since the experiments of Ref.~\cite{Sonier:97a}, strong field 
and temperature dependences of the 
vortex-core size have been found by $\mu$SR in a variety of 
superconductors. Through comparison of the results with theoretical
models and experiments that are directly sensitive to quasiparticles
properties, a good understanding of many of the $\mu$SR experiments 
has been achieved. In $s$-wave superconductors, it is now well
established that the vortex core size depends on both the thermal 
occupancy of the bound quasiparticle core states and the overlap of the 
corresponding quasiparticle wave functions with those of nearest-neighbour 
vortices. However, in exotic systems such as high-$T_c$ superconductors, 
where localized cores states may be absent, there is currently
insufficient experimental information to make similar definitive statements.
On the other hand, recent $\mu$SR studies of the vortex cores in 
underdoped high-$T_c$ superconductors have shed new light on the 
ground state that emerges when superconductivity is suppressed.
Combining information obtained from $\mu$SR experiments in zero
and nonzero magnetic fields, the latest results support a picture
of closely competing superconducting and magnetic ground states.          

\section{Muon spin rotation ($\mu$SR)}

\subsection{General description}

The acronym $\mu$SR dates back to 1974, and stands for either 
`muon spin relaxation, rotation' or `resonance'. 
These three terms describe different uses of the magnetic moment
of a muon ($\mu^-$ or $\mu^+$) to probe matter. While the principle
aspects of the technique are analogous to nuclear magnetic resonance
(NMR), there are many important differences. Only some of these are
touched upon in this review. 

The primary use and strength of
$\mu$SR is its unmatched sensitivity to internal magnetism. 
Central to the $\mu$SR method is the use of 
a nearly 100~\% spin-polarized muon beam, naturally generated from 
the weak interaction decay of pions. This is a great advantage over
conventional NMR, which relies on thermal equilibrium nuclear spin polarization
in a large magnetic field. Zero-field (ZF) $\mu$SR is routinely used 
to study small internal magnetic fields of natural origin. In contrast
to neutron scattering, the information provided by $\mu$SR is integrated 
over reciprocal space, which makes it ideal for studies of short-range 
magnetic correlations or disordered magnetism. 
The magnetic moment of the muon is 3.18 
times larger than that of the proton, making it even more sensitive 
to magnetism than NMR. Although generally a nuisance in experiments, 
$\mu$SR even detects the dipolar fields of nuclear moments. In fact,
magnetic fields as small as $\sim \! 0.1$~G are detectable---although
it is important to emphasize that this refers to the local field at
the muon stopping {\it site}.

\subsection{Transverse-field $\mu$SR}  

\begin{figure}
\centering
\includegraphics[width=9.0cm]{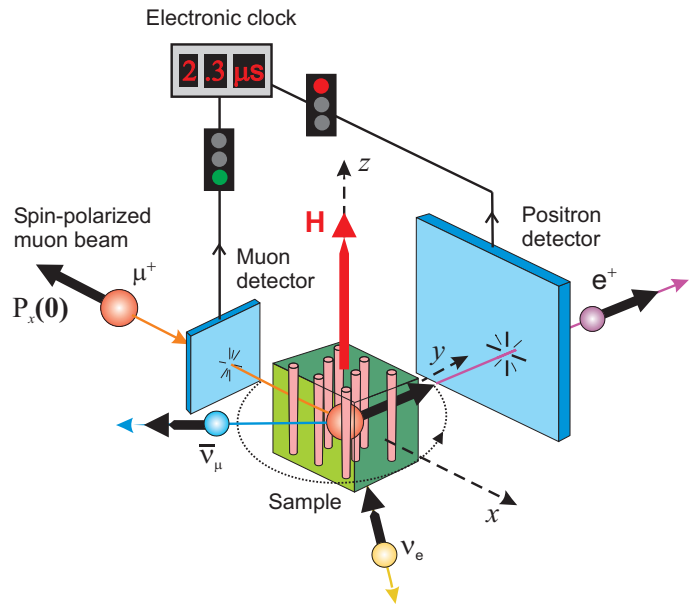}
\caption{Schematic of the arrangement for 
a TF-$\mu$SR experiment on a type-II superconductor in the vortex state.
The muon spin Larmor precesses about the local magnetic 
field $B$ at its stopping site in the sample, and subsequently undergoes
the three-body decay $\mu^+ \rightarrow e^+ + \nu_e + \bar{\nu}_{\mu}$.
The time evolution of the muon spin polarization $P_x(t)$ is accurately
determined by detection of the decay positrons from $\sim \! 10^6$ 
muons implanted one at a time.}   
\label{fig1}
\end{figure}

The internal magnetic field distribution $n(B)$ of a type-II 
superconductor in the vortex state is measured by the so-called
`transverse-field' muon spin rotation (TF-$\mu$SR) method. The geometry
of a TF-$\mu$SR experiment is shown in figure \ref{fig1}. The external
magnetic field $H$ is applied {\it transverse} to the direction of the
initial muon spin polarization $P_x(0)$, which defines the $x$-axis.
In high-$T_c$ superconductors, the positive muon ($\mu^+$) forms an 
$\sim \! 1$~\AA~ bond with an oxygen atom \cite{Dawson:88,Brewer:90}, 
but in general the muon will
stop at an interstitial site in the sample. There the  
muon spin precesses about the local magnetic 
field {\bf B}({\bf r}) in a plane perpendicular to the local field axis.
The muon subsequently decays, emitting a fast positron. The angular
dependence of the decay probability of the muon is given by
\begin{equation}
W(E, \theta) = 1 + a(E) \cos(\theta) \, ,
\end{equation}
where $E$ is the kinetic energy of the decay positron, $\theta$ is
the angle between the directions of the muon spin and the emitted 
positron, and $a(E)$ is an asymmetry factor. When all positron energies
are sampled with equal probability, the asymmetry factor has the
value $a \! = \! 1/3$. The statistical average direction of
the muon spin polarization is obtained by measuring the anisotropic
angular distribution of decay positrons from an ensemble of implanted muons.   

The $\mu$SR signal obtained by the detection of the decay positrons 
is given by
\begin{equation}
A(t)  = a_0 P_x(t) \, , 
\end{equation}
where $A(t)$ is the $\mu$SR `asymmetry' spectrum, 
$a_0$ is the asymmetry maximum, 
and $P_x(t)$ is the time evolution of the muon spin polarization
\begin{equation}
P_x(t) = \int_0^{\infty} n(B) \cos(\gamma_\mu B t + \phi) dB \, .
\label{eq:polarization}  
\end{equation}
Here $\gamma_\mu \! = \! 0.0852$~$\mu$s$^{-1}$~G$^{-1}$ is the muon 
gyromagnetic ratio, $\phi$ is a phase constant, and
\begin{equation}
n(B^{\prime}) = \langle \delta [ B^{\prime} - B({\bf r})] \rangle \, ,
\end{equation}
is the probabilty of finding a local magnetic field $B$ in the $z$-direction 
at a position {\bf r} in the $x$-$y$ plane.

\subsection{Application to studies of the vortex state}

The internal magnetic field distribution $n(B)$ measured by $\mu$SR is
similar to what one can measure with NMR. 
However, because the muon has a spin equal to 1/2, it does not possess 
an electric quadrupole moment and hence is a pure magnetic probe.
In contrast, $n(B)$ measured by NMR often includes quadrupolar broadening
and in some cases closely spaced quadrupolar satellites. 
At its stopping site, a given muon {\it randomly} 
samples $n(B)$. This is because the intervortex spacing 
is typically several orders of magnitude larger than the atomic lattice
spacing. For this reason it is often not necessary to know precisely
where the muon stops in the sample.

\begin{figure}
\includegraphics[width=10.5cm]{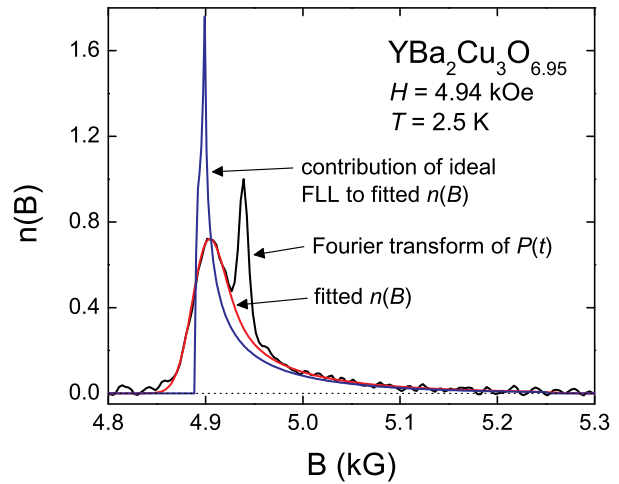}
\caption{Fourier transform of the TF-$\mu$SR signal from
single crystal YBa$_2$Cu$_3$O$_{6.95}$ at $T \! = \! 2.5$~K 
and $H \! = \! 4.94$~kOe (black curve) computed
using a Gaussian apodization function with a width of 
3~$\mu$s$^{-1}$. The peak at $B \! \approx \! 4.94$~kG is 
a background signal coming from muons stopping outside the
sample. Also shown is the internal magnetic field distribution
$n(B)$ (red curve) obtained from fits to the $\mu$SR precession 
signal. The fit assumes $n(B)$ of an ideal FLL (blue curve)
generated from a Ginzburg-Landau model described later, and
additional broadening by FLL disorder and nuclear dipole moments.} 
\label{fig2}

\end{figure}

Although the $\mu$SR signal is recorded in the time domain, a fairly 
accurate visual illustration of $n(B)$ is provided by the Fourier
transform of $P(t)$---often called the `$\mu$SR line shape'.
An example of a $\mu$SR line shape for YBa$_2$Cu$_3$O$_{6.95}$ is
shown in figure~\ref{fig2}. Included in this figure is $n(B)$
of equation~(\ref{eq:polarization}) obtained from a fit to the
asymmetry spectrum, the details of which are described later in 
this article. Despite the Gaussian
apodization used in generating the Fourier transform, the $\mu$SR
line shape closely resembles $n(B)$ obtained from the fit to the
$\mu$SR signal in the time domain. Note that $n(B)$
differs somewhat from that expected for an ideal FLL, because of 
FLL disorder and the contribution from nuclear dipoles.
   
\begin{figure}
\includegraphics[width=8.0cm]{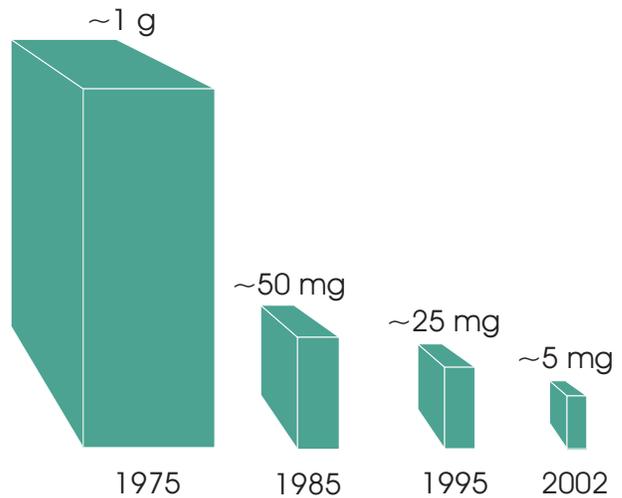}
\caption{Evolution of the minimum sample size
required for $\mu$SR studies of the vortex state.}   
\label{fig3}
\end{figure}
 
As shown in figure~\ref{fig3}, the minimum size of the sample required for 
$\mu$SR studies has greatly decreased over the years.
Today's capability to probe smaller samples comes from several key 
developments. Like neutron scattering, the quality of $\mu$SR spectra
are limited by counting times and background contributions.
Shorter data collection times and smaller background signals have
been achieved through the availibility of higher luminosity muon beam lines, 
which allow more of the beam to be focused on the sample.
Major advancements in $\mu$SR spectrometer design 
\cite{Schneider:93,Chakhalian:97}
have also greatly helped in suppressing the background signal arising fom muons
that miss the sample and stop elsewhere. In many setups this is 
achieved by placement of the muon 
and positron counters within the same cryogenic environment as the
sample. This tighter geometry makes it easier to restrict the {\it good} 
decay events to those associated with muons stopping in the sample.           
The background peak that contributes to the $\mu$SR line shape in 
figure~\ref{fig2} constitutes $\sim \! 12$~\% of the total signal,
and is accounted for by adding a polarization function 
$\exp(-\sigma_{\rm bg}^2t^2/2)\cos(\gamma_\mu B_{\rm bg}t + \phi_{\rm bg})$ 
with 12~\% amplitude to equation~(\ref{eq:polarization}). 
This $\mu$SR line shape was
recorded in 1992. As shown later, today this same measurement 
can be done with virtually no background signal. However, it is 
often helpful in the anaylsis of such data to have some background, 
as it provides an accurate determination of the average magnetic field 
at the sample position.    

\section{The muon depolarization rate}

In the London model the local magnetic field for a perfect FLL is
\begin{equation}
B({\bf r}) = B_0 \sum_{ {\bf G}}
\frac{e^{-i {\bf G} \cdot {\bf r}}}{1 + \lambda^2 G^2} \, ,
\label{eq:London}
\end{equation}
where the sum is over reciprocal lattice vectors {\bf G}
of the FLL.
This equation is considered valid provided 
$\lambda \! \gg \! \xi$ and $B \! \ll \! B_{c2}$, where $\xi$ is the superconducting 
coherence length and $B_{c2}$ is the upper critical field. 
The second moment of the internal magnetic field distribution $n(B)$ corresponding to 
equation (\ref{eq:London}) is \cite{Brandt:88b}
\begin{equation}
\langle (\Delta B)^2 \rangle = \langle ( B({\bf r}) - \langle B \rangle)^2 
\rangle =  \eta \Phi_0^2 \lambda^{-4} \, ,
\label{eq:moment}
\end{equation}
where $\langle B \rangle$ is the average of $n(B)$, 
$\Phi_0 \! = \! hc/2e \! = \! 2.07 \! \times \! 10^{-15}$~T-m$^2$, and
$\eta \! = \! 0.00371$ and $\eta \! = \! 0.003868$ for a
hexagonal and square FLL, respectively \cite{Brandt:91}. In the London model
$\langle (\Delta B)^2 \rangle$ is independent of the applied field.
In high-$T_c$ superconductors, where the effective mass of the charge carriers
is larger in the direction perpendicular to the CuO$_2$ layers compared to
that parallel to the CuO$_2$ layers, $\lambda$ is highly anisotropic.
If the field is applied along the $\hat{c}$-axis, then 
$\lambda \! = \! (\lambda_a \lambda_b)^{1/2} \! \equiv \! \lambda_{ab}$,
where $\lambda_{ab}$ is the in-plane magnetic penetration depth describing
screening currents flowing in the $\hat{a}$-$\hat{b}$ plane around a vortex.
Applying the field perpendicular to the $\hat{c}$-axis results in the formation
of Josephson vortices sandwiched between CuO$_2$ layers.
In this case the screening currents flow in the $\hat{a}$-$\hat{c}$ or $\hat{b}$-$\hat{c}$ 
plane, and $\lambda \! = \! (\lambda_a \lambda_c)^{1/2} \! \equiv \! \lambda_{ac}$
or $\lambda \! = \! (\lambda_b \lambda_c)^{1/2} \! \equiv \! \lambda_{bc}$, respectively.
The out-of-plane penetration depth $\lambda_{c}$ is typically 
much greater than 5000~\AA,
so that $\langle (\Delta B)^2 \rangle$ is significantly reduced in this geometry.
For such large values of the magnetic penetration depth, the width of the
$\mu$SR line shape in the vortex state becomes comparable to that in the
normal state, and this prevents an accurate determination of the temperature
dependence of $\lambda$.
The absolute values of $\lambda_a$, $\lambda_b$ and $\lambda_c$ cannot be
isolated by $\mu$SR without a priori knowledge of the mass anisotropies.
For example, if it is known that $\lambda_b \! = \! \lambda_a$,
then $\lambda_a$ can be determined from a measurement of $\lambda_{ab}$, and 
$\lambda_c$ can be determined from a subsequent measurement of $\lambda_{ac}$. This
has been done for the heavy-fermion superconductor UPt$_3$ \cite{Yaouanc:98}. 
          
For polycrystalline samples an often used approximation is to 
assume $n(B)$ is a simple Gaussian distribution of fields of
width $\sigma$
\begin{equation}
n(B) = \frac{\gamma_\mu}{\sqrt{2 \pi} \sigma} \exp \left( -\frac{1}{2} 
\frac{\gamma_\mu^2 B^2}{\sigma^2} \right) \, ,
\end{equation}
in which case equation (\ref{eq:polarization}) has the following form
\begin{equation}
P_x(t) = \exp(- \sigma^2 t^2/2)\cos(\gamma_\mu \langle B \rangle t + \phi) \, .
\end{equation}
The width $\sigma$ is thus equivalent to the damping rate of the asymmetry spectrum,
called the `muon depolarization rate'. There is generally a temperature-independent 
contribution to $\sigma$ from randomly oriented nuclear dipole fields, 
which can easily be determined from measurements above $T_c$.
The resultant measured value of $\sigma$ can then be 
used to obtain the second moment of the Gaussian distribution $n(B)$
\begin{equation}
\langle (\Delta B)^2 \rangle =  \sigma^2/\gamma_\mu^2 \, .
\label{eq:momentGauss}
\end{equation}        
By equating equation (\ref{eq:moment}) to equation (\ref{eq:momentGauss}),
one gets an estimate of $\lambda$
\begin{equation}
\sigma =  \eta \gamma_\mu \Phi_0 \lambda^{-2} \, .
\label{eq:sigma}
\end{equation}
For large anisotropy 
$\gamma \! \equiv \! (m_c/m_{ab})^{1/2} \! = \! \lambda_c/ \lambda_{ab} \! > \! 5$
the effective penetration depth $\lambda$ is primarily determined by
$\lambda_{ab}$, such that
\begin{equation}
\lambda \! \simeq \! f \lambda_{ab} \, ,
\label{eq:anisotropy}
\end{equation}
where $f \! = \! 1.23$ \cite{Barford:88}. For $\gamma \! = \! 1$ to 5, $f$
varies from 1 to 1.23.

Early $\mu$SR measurements of $\sigma$ were used to deduce information 
about $1/\lambda_{ab}^2$ using equations~(\ref{eq:sigma}) and (\ref{eq:anisotropy}). 
From observations of a
weak temperature dependence for $\sigma$ at low $T$ it was concluded that
the pairing state symmetry of high-$T_c$ superconductors is $s$-wave
\cite{Harshman:87,Uemura:88,Harshman:89,Pumpin:90}.
A linear scaling of $T_c$ with $\sigma$ was also found (the so-called `Uemura plot'), 
which has been interpreted via equation (\ref{eq:sigma}) to imply 
$T_c \! \propto \! 1/\lambda_{ab}^2$, which in turn is proportional
to the superfluid density $n_s$ \cite{Uemura:89,Uemura:91}.
While the dependences of $\sigma$ for newly discovered superconductors
continue to be reported in the literature, the assumed relation 
$\sigma \! \propto \! 1/\lambda_{ab}^2$ is not precisely correct.  
One reason is that the use of equation~(\ref{eq:sigma}) assumes one is aware 
of and understands all contributions to the measured value of $\sigma$. 
Brandt \cite{Brandt:88b} has shown that for
straight rigid vortex lines, $\sigma$ is increased by  
both pinning-induced random displacements of the vortices
from their ideal position in the FLL and thermal fluctuations of the vortex lines. 
On the other hand, the layered nature of the high-$T_c$ superconductors means that
the vortices can be highly flexible. In this case segments of a flux line are susceptible 
to pinning or thermal-induced displacements, which reduce $\sigma$.
In either situation the size of the effect will be dependent on temperature and
the applied magnetic field. There may also be contributions to $\sigma$ from static 
electronic moments. This is likely to be the case in rare-earth and lightly-doped
cuprate superconductors. Even if identified, these contributions to $\sigma$ 
cannot be reliably 
separated from the depolarization rate associated with $\lambda_{ab}$.   

A second consideration is that 
Yaouanc, Dalmas de R\'{e}otier and Brandt \cite{Yaouanc:97} 
have shown that the second moment due to an ideal hexagonal FLL is 
more accurately given by
\begin{equation}
\langle (\Delta B)^2 \rangle = 0.00371 \Phi_0^2 \lambda^{-4} f_v(b),
\label{eqtn:Yaouanc}
\end{equation}
where $b \! = \! B/B_{c2}$ is the reduced field and $f_v(b)$ is a universal function 
that accounts for the finite size of the vortex cores.
It is often assumed that the vortex cores of high-$T_c$ superconductors
are small ({\it i.e.} $\xi_{ab} \! \approx \! 15$~\AA), so that little error 
is introduced in neglecting them. However, measurements by other 
techniques indicate that the vortex core radius can greatly exceed 
15~\AA~ in underdoped and overdoped samples \cite{Ando:02,Wen:03,Wang:03}. 
As explained in Refs.~\cite{Yaouanc:97,Brandt:03}, $f_v$ is strongly dependent on 
$b$. Moreover, Brandt \cite{Brandt:03} has shown that the often used 
approximation $f_v \! = \! 1$ is really only satisfactory over a narrow field range 
and only for $\kappa \! \geq \! 70$. The latter condition is not 
always satisfied in high-$T_c$ superconductors.

\begin{figure}
\includegraphics[width=10.5cm]{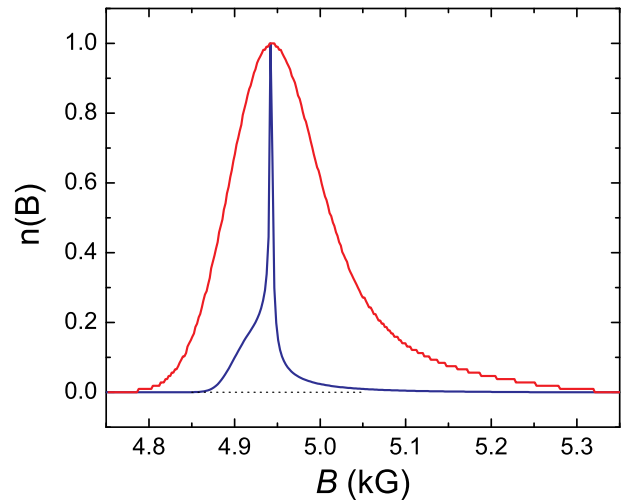}
\caption{Theoretical field distributions for an 
{\it unoriented polycrystalline} sample calculated from the 
`Kossler model' for the powder average of an anisotropic
superconductor \cite{Celio:88} and including FLL disorder. 
The parameters used to generate
the blue curve come from the fit of {\it single crystal} 
YBa$_2$Cu$_3$O$_{6.95}$ in figure~\ref{fig2}.
The red curve is calculated with a higher degree of FLL 
disorder ({\it i.e.} $\sigma_d$ of equation~(\ref{eq:disorder}) is
three times larger), and is closer to
a Gaussian distribution.
(Note that the two curves for $n(B)$ are not normalized with
respect to each other.)} 
\label{fig4}
\end{figure}

A third problem is the assumption that $n(B)$ is a Gaussian distribution.
In a polycrystalline sample $n(B)$ is an average over all orientations 
\cite{Brandt:88b,Celio:88}. An example of what $n(B)$ should look like in
an unoriented powder sample is shown in figure~\ref{fig4}.
The fact that many of the early $\mu$SR line shapes from measurements 
on polycrystalline samples of high-$T_c$ superconductors looked Gaussian-like, 
suggests that the samples were of poor quality with a high-degree of FLL
disorder.
 
Given these issues, the dependences of $\sigma$ should be viewed as
providing the correct qualitative trends of $1/\lambda^2$ only. 
Indeed the Uemura plot was an important achievement, as it established
for the first time a qualitative relationship between $T_c$ and the 
superfluid density.
However, the current state of affairs regarding the search for a 
microscopic theory of high-$T_c$ superconductivity is such that 
precise quantitative relations are desired. In recent years, experimental
studies of $1/\lambda^2$ have shown that $T_c$ is a sublinear function 
of $n_s$ in the underdoped regime \cite{Pereg:04,Zuev:05,Broun:05}
and that there is no single universal Uemura relation for
high-$T_c$ cuprates \cite{Tallon:03}. Likewise, experiments on single 
crystals \cite{Hardy:93}, including $\mu$SR \cite{Sonier:94,Luke:97}, 
have established a limiting low-temperature linear $T$ dependence of 
$\lambda_{ab}$, that is not evident in the earlier measurements of
$\sigma$. This behaviour is consistent with a $d_{x^2-y^2}$-wave 
superconducting order parameter, which has been clearly identified
by phase-sensitive techniques \cite{Harlingen:95}. 

\section{Full line shape analysis}
To extract values of $\lambda$ and the vortex core size from 
TF-$\mu$SR measurements, one must assume a model for the internal magnetic 
field distribution $n(B)$. This in fact is the largest source of uncertainty
in these kinds of measurements.
To date, the analysis of $\mu$SR measurements in the vortex state have
primarily relied on models based on either the phenomenological London or 
Ginzburg-Landau (GL) theories. The reason is that in
some limiting cases, analytical models for
the spatial field profile $B({\bf r})$ exist for these theories.
While the analytical London and GL models provide very good fits to
$\mu$SR measurements of the FLL in real materials,
neither are generally valid, and considerable effort 
has been devoted in recent years to understanding the meaning of 
the fitted parameters $\lambda$ and $\xi$. In particular, it is
important to consider the following cautionary statements, which are
discussed in further detail in the following sections:
\begin{itemize}
\item Generally speaking, {\bf the fitted parameter $\lambda$ is not 
a measure of the magnetic penetration depth.} This is because $\lambda$ 
is influenced by deviations of the assumed theoretical model for $n(B)$ 
from the actual field distribution of the FLL in the sample. This point was made in 
Refs.~\cite{Yaouanc:97,Amin:98,Sonier:99b,Amin:00,Sonier:04c,Callaghan:05,Laulajainen:06,
Laiho:06,Laiho:07,Landau:07}.     
For this reason, the magnetic penetration depth actually corresponds to the
extrapolated $H \! \rightarrow \! 0$ value of $\lambda$.

\item Likewise, {\bf the fitted parameter $\xi$ is generally not a 
measure of the superconducting coherence length.} This point is explained
in detail in Ref.~\cite{Sonier:04c}. Instead $\xi$ qualitatively tracks
changes in the vortex core size that have nothing to do with
the size of the Cooper pairs.   
\end{itemize} 

On the theoretical front, great strides have been made in numerically 
calculating $n(B)$ from the microscopic equations. The predictions of the microscopic theory
provide one means of determining the accuracy of the phenomenological approach to 
modelling $\mu$SR measurements.         

\subsection{Finite core size}

\subsubsection{Modified London models.}
The London model does not account for the spatial dependence of the superconducting
order parameter and consequently equation ({\ref{eq:London}) breaks down at distances
on the order of $\xi$ from the vortex core centre---{\it i.e.} $B(r)$ diverges as
$r \! \rightarrow \! 0$. To correct this, the ${\bf G}$ sum can be truncated
by multiplying each term in equation ({\ref{eq:London}) by a cutoff function
$F(G)$. It is important to stress that the appropriate form of $F(G)$ depends 
on the precise spatial dependence of the order parameter 
in the the vortex core region, and this in general depends on temperature 
and magnetic field. 

Analytical expressions for $F(G)$ 
exist only for certain limiting cases. Some time ago Brandt suggested the use of a 
smooth Gaussian cutoff factor
\begin{equation}
F(G) \! = \! \exp(- \alpha G^2 \xi^2) \, .
\end{equation}
If there is no dependence of the superconducting coherence length $\xi$ 
on temperature and magnetic field, then changes
in the spatial dependence of the order parameter around a vortex
correspond to changes in $\alpha$. By solving the GL equations, Brandt 
determined that $\alpha \! = \! 1/2$ at fields near $B_{c2}$ 
\cite{Brandt:88b,Brandt:77,Brandt:88a}, and arbitarily determined that
$\alpha \! \approx \! 2$ at fields immediately above 
$B_{c1}$ \cite{Brandt:92}.
More accurate values of $\alpha$ have since been obtained from precision
solutions of the GL equations. For an isolated vortex in an 
isotropic extreme ($\kappa \! \gg \! 1$) $s$-wave superconductor, 
$\alpha$ is found to decrease smoothly from $\alpha \! = \! 1$ 
at $B_{c1}$ to $\alpha \! \approx \! 0.2$ at $B_{c2}$ \cite{Oliveira:98}.
Even so, it is rather presumptuous to incorporate the field dependence
of $\alpha$ into the analysis of $\mu$SR spectra, as the vortices 
in real materials are not isolated. Furthermore, the 
additional effects of temperature on the spatial dependence of the 
order parameter in the vortex core region are not accounted for in 
calculations of $\alpha(B)$. For example, 
Laiho {\it et al.} \cite{Laiho:06,Laiho:05} have shown by
comparison to solutions of the quasiclassical Eilenberger equations
for a $d_{x^2-y^2}$-wave superconductor that $\alpha$ is 
temperature dependent. More recently they have determined the
magnetic and temperature dependences of the cutoff function
used in the London model by numerically solving the quasiclassical Eilenberger
equations for the vortex state of an $s$-wave superconductor \cite{Laiho:07}.
These calculations show that the London model with the proper
cutoff function provides a reasonable description of the field 
distribution of the FLL in real superconductors.
       
Generally, $\alpha$ is fixed when
using the modified London model to fit $\mu$SR spectra, so that
the temperature and magnetic field dependences of the order 
parameter are indicated by variations in the fitted value of $\xi$. 
Such a procedure has been used to account for the finite size of the 
vortex cores in a number of $\mu$SR studies of the vortex state in
type-II superconductors 
\cite{Sonier:94,Riseman:95,Sonier:97c,Kadono:01,Ohishi:02,Price:02}.
Unfortunately, none of these studies were done at sufficiently
low enough temperature to make a direct comparison of the field
dependence of $\xi$ with the theoretical predictions for 
$\alpha(B)$ at $T \! = \! 0$~K.

\subsubsection{Ginzburg-Landau models.}
In recent years, modified London models for $B({\bf r})$ have been 
abandoned by some in favour of models based on GL theory. The appealing aspect
of the GL models is that the spatial variation of the order parameter 
is naturally built into the theory. The drawback is that GL theory 
assumes that the order parameter varies slowly in space and is strictly 
valid only near $T_c$. Despite these limitations, GL theory has proven
to be highly successful in describing variations of $n(B)$ as measured
by $\mu$SR, yielding accurate quantitative values of $\lambda$ and
$\xi$ in certain cases. As is the case in using modified London models, 
the key is to be careful with the interpretation of the fitted values.

The GL equations for the ideal Abrikosov vortex lattice can be solved
by a variational method \cite{Clem:75}. At low reduced fields 
$b \! = \! B/B_{c2} \! \ll \! 1$ and for $\kappa \! \gg \! 1$, an
excellent analytical approximation to the spatial field profile in
GL theory is \cite{Yaouanc:97}
\begin{equation}
B({\bf r}) = B_0 (1-b^4) \sum_{ {\bf G}}
\frac{e^{-i {\bf G} \cdot {\bf r}} \, \, F(G)}{\lambda^2 G^2} \, ,
\label{eq:Yaouanc}
\end{equation}
where $F(G) \! = \! u \, K_1(u)$,
$u^2 \! = \! 2 \xi^2 G^2 (1 + b^4)[1-2b(1 - b)^2]$, and $K_1(u)$ is a 
modified Bessel function. Note the cutoff function $F(G)$ depends
on the local internal magnetic field $B$. 

Brandt later introduced an interative 
procedure for solving the GL equations for arbitrary $b$, $\kappa$ and
vortex-lattice symmetry \cite{Brandt:97}. While more generally applicable,
fitting $\mu$SR spectra with this iterative method requires a tremendous
amount of computer time. Fortunately, it does not seem that this is
necessary. In a recent $\mu$SR study of the
low-$\kappa$ type-II superconductor V, qualitatively similar results
were obtained from analyses with the variational and iterative
solutions of the GL equations \cite{Laulajainen:06}.  

\begin{figure*}
\includegraphics[width=16.0cm]{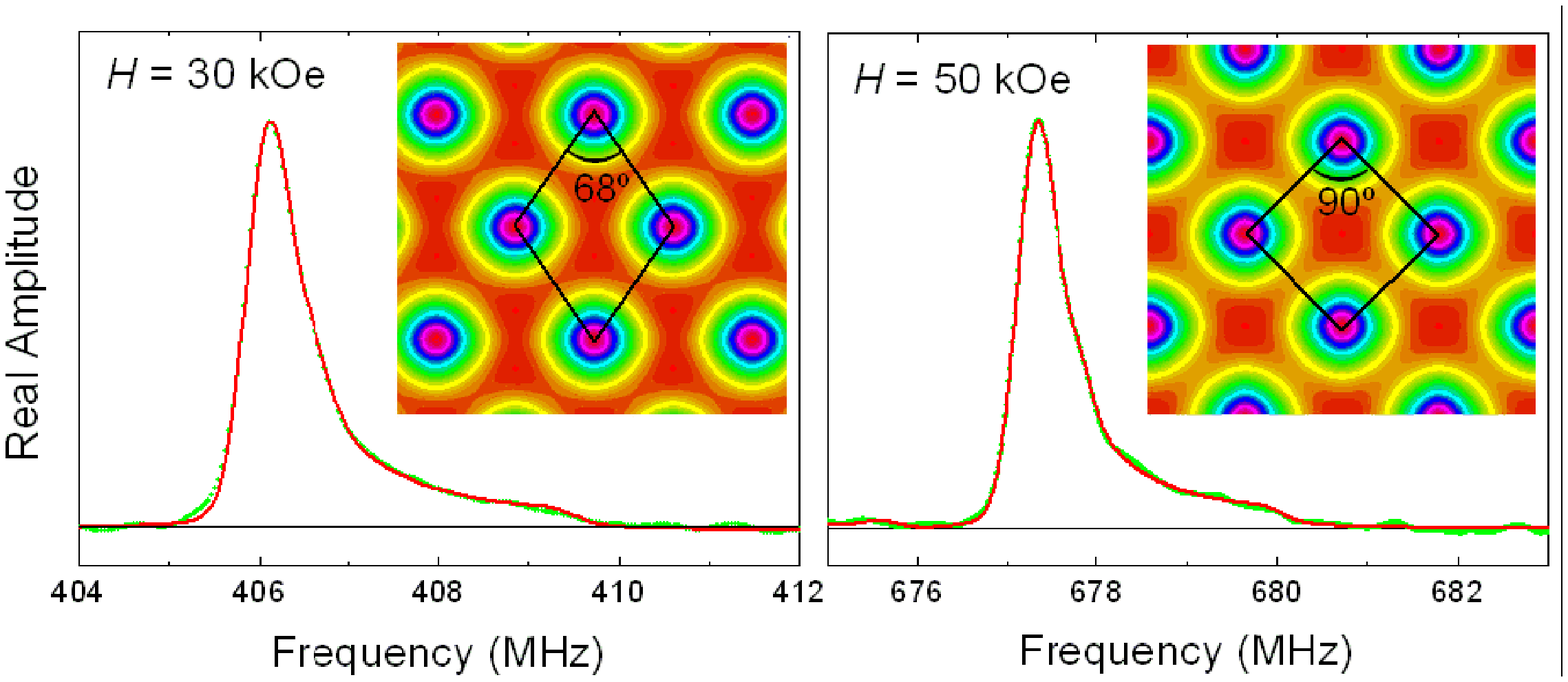}
\caption{Fourier transforms of the TF-$\mu$SR signal from
single crystal V$_3$Si at $T \! = \! 3.8$~K for
applied magnetic fields of $H \! = \! 30$~kOe and 
$H \! = \! 50$~kOe directed parallel to the [001]
axis (from data in Ref.~\cite{Sonier:04a}). 
The solid red curve in each panel is a Fourier 
transform of the fit in the time domain (Note: because
of apodization, this is slightly different than the fitted
$n(B)$, an example of which was shown in figure~\ref{fig2}). 
Also shown are contour plots of the field profile $B({\bf r})$ 
obtained from the fits to a GL model. The results are in
good agreement with STS measurements on V$_3$Si 
\cite{Sosolik:03}.}       
\label{fig5}
\end{figure*}

\subsubsection{Vortex symmetry.} 
In the above models the individiual vortices  are assumed 
to have circular symmetry. However, this is not the case in real type-II 
superconductors with anisotropic Fermi surfaces and/or superconducting energy gaps
\cite{Takanaka:71,Ichioka:96,Kogan:97a,Kogan:97b,Nakai:02}. At
magnetic fields where there is an appreciable degree of overlap 
of the vortices, the FLL adopts the symmetry of the individual flux
lines. In cases where the deviation from circular symmetry is extreme, 
the symmetry of the FLL can be accounted for in the model for $B({\bf r})$.
This has been successfully done in $\mu$SR studies of YNi$_2$B$_2$C \cite{Ohishi:02}, 
LuNi$_2$B$_2$C \cite{Price:02}, Sr$_2$RuO$_4$ \cite{Luke:99}, 
V$_3$Si \cite{Sonier:04a} and Nb$_3$Sn \cite{Kadono:06}.
In the case of V$_3$Si, it was possible to
extract from the $\mu$SR measurements a gradual hexagonal-to-square
transformation of the FLL symmetry in good agreement 
with STS imaging experiments. A couple of measurements from this study
are shown in figure~\ref{fig5}. 

In Refs.~\cite{Ohishi:02,Price:02,Sonier:04a}, the theoretical models
used for analysis of the TF-$\mu$SR spectra are based on London theory
\cite{Kogan:97a,Kogan:97b}. London models that
account for the fourfold symmetry of the $d_{x^2-y^2}$-wave 
order parameter in cuprate superconductors have also been developed
\cite{Affleck:96,Amin:98}. Anisotropy may also be incorporated into
analytical GL models. The TF-$\mu$SR study of Sr$_2$RuO$_4$ \cite{Luke:99} 
employed a GL model that accounts for $p$-wave symmetry and Fermi surface 
anisotropy \cite{Agterberg:98,Heeb:99}. Since all of the theoretical models 
accounting for anisotropy introduce additional fitting parameters, they
are seldom used in $\mu$SR studies. When anisotropy is not included
in the model for $B({\bf r})$, the fitted values of $\lambda$ and $\xi$ 
are angle-averaged length scales.
 
\subsection{Vortex Lattice Disorder}
\subsubsection{3D vortex lines.} 
The effects of random pinning and thermal fluctuations of the vortices 
on $n(B)$ depend very much on the rigidity of the vortex lines.
In general, reliable information on the length scales $\lambda$ and 
$\xi$ is obtained only when the vortices are fairly straight and 
parallel. In this case, pinning or fluctuation induced displacements of 
the vortices from their positions in the perfect FLL increase the width of 
$n(B)$. The effect is nearly equivalent to smearing
$n(B)$ by convolution with a Gaussian distribution of fields \cite{Brandt:88a}
\begin{equation}
n(B) = \int \frac{1}{\sqrt{2 \pi} \sigma_d}\exp\left[ - \frac{1}{2} 
\left( \frac{B - B^{\prime}}{\sigma_d} \right) ^2 \right] 
n(B^{\prime}) dB^{\prime} \, ,
\label{eq:disorder}
\end{equation}
where $\sigma_d$ is related to the root mean square displacement 
$\langle u_l^2 \rangle^{1/2}$ of a vortex line about its average 
position. From $\mu$SR studies on numerous superconductors with 3D-like
vortices it has been determined that $\sigma_d \! \propto \! B/\lambda_{ab}^2$.
This simply means that stronger overlap of neighbouring vortices reduces the
degree of disorder. 

The above treatment is a reasonable approximation of the weak
FLL disorder in the 3D `Bragg-glass' phase of a type-II superconductor, 
in which quasi-long-range translational order and perfect topological order 
are preserved \cite{Giamarchi:95}. Note that a more accurate 
expression for the variance of the FLL field distribution in the Bragg-glass 
phase has been derived \cite{Dalmas:04}, but has yet to be used for the
analysis of $\mu$SR measurements. 
More recently, it has been shown that when disorder
is strong enough to produce a 3D `vortex-glass' phase, in which the topological
order of the FLL is not retained, the $\mu$SR line shape is slightly skewed in
the opposite way with a `tail' on the low-field side 
\cite{Divakar:04,Menon:06}. In this situation, the disorder cannot be
handled in a trivial way.  

\subsubsection{2D pancake vortices.}  
The vortex lines in layered superconductors are generally considered to
be comprised
of coupled 2D `pancake' vortices. When the interlayer coupling is weak,
the vortex lines are very {\it soft}, and pinning and/or thermal fluctuations 
of the pancake vortices shorten the high-field tail of $n(B)$ 
\cite{Brandt:91,Koshelev:96,Menon:99}.

There has been some debate on how disorder of the FLL should be treated 
in $\mu$SR studies of high-$T_c$ superconductors. 
An early $\mu$SR study by Harshman {\it et al.}
\cite{Harshman:93} arrived at the conclusion that 
point distortions of the vortex lines was
significant in highly anisotropic Bi$_2$Sr$_2$CaCu$_2$O$_{8 + \delta}$, as evidenced
by narrow symmetric $\mu$SR line shapes at low temperatures and high magnetic fields.
On the other hand, under similar conditions broad asymmetric $\mu$SR line shapes 
were observed for YBa$_2$Cu$_3$O$_y$, consistent with vortex
lines that do not wander significantly along their length.
The dimensionality of the vortex lines is dependent on the ratio 
$\gamma s/ \lambda_{ab}$, where $\gamma$ is the mass anisotropy and $s$ is
the spacing between CuO$_2$ planes. The observations reported in 
Ref.~\cite{Harshman:93} are consistent with the fact that $\gamma$ is nearly 
two orders of magnitude smaller in optimally doped YBa$_2$Cu$_3$O$_y$  
than in Bi$_{2}$Sr$_{2}$CaCu$_2$O$_{8+\delta}$.
Thus in high-$T_c$ cuprates the picture of weakly coupled 2D 
pancake vortices should be viewed as a limiting case. 

Recently, Harshman {\it et al.} \cite{Harshman:04} have asserted that 
point distortions of the vortex lines in {\it ultra-clean} samples of
YBa$_2$Cu$_3$O$_y$ are as important as in Bi$_{2}$Sr$_{2}$CaCu$_2$O$_{8+\delta}$.
By fitting the total second moment 
of the $\mu$SR line shape for fully doped YBa$_2$Cu$_3$O$_7$ to a phenomenological
model containing two independent parameters for pinning-induced distortions of the FLL, 
Harshman {\it et al.} argued that $\lambda_{ab}(T, H)$ is consistent with 
$s$-wave superconductivity. However, there are several problems with the analysis of the 
$\mu$SR data in Ref.~\cite{Harshman:04}. Most notably, the second moment of $n(B)$
is assumed to be given by equation (\ref{eq:moment}), which does not account for
the finite size of the vortex cores. Also, both the determined value 
$\kappa \! = \lambda_{ab}/\xi_{ab}\! = \! 43.8 \! \pm 1.8$ and
the lowest field ($B \! = \! 0.05$~T) considered in Ref.~\cite{Harshman:04}
are too small for equation (\ref{eq:moment}) to apply. 
The smaller value of the second moment found at $B \! = \! 0.05$~T 
in Fig.~2 of Ref.~\cite{Harshman:04} is in fact expected, because at this field
$\langle (\Delta B)^2 \rangle$ is more appropriately described by equation (12) 
of Ref.~\cite{Brandt:03}

\begin{equation}
\langle (\Delta B)^2 \rangle = \frac{b \kappa^2}{8 \pi^2} \frac{\Phi_0^2}{\lambda^4} \, .
\end{equation}

Although the vortex lines in YBa$_2$Cu$_3$O$_y$ and La$_{2-x}$Sr$_x$CuO$_4$ are
certainly more rigid than in Bi$_{2}$Sr$_{2}$CaCu$_2$O$_{8+\delta}$, they do
soften in underdoped samples due to an increase in anisotropy $\gamma$. 
Experiments on oxygen-deficient YBa$_2$Cu$_3$O$_y$ thin films
\cite{Deak:95,Sefrioui:99} indicate that for applied magnetic fields currently    
attainable in a TF-$\mu$SR experiment ($H \! = \! 80$~kOe), 
the FLL remains 3D at low $T$,
except perhaps in lightly doped samples. However, the dimensionality
of the vortex system depends on the degree of sample disorder, 
which can cause pinning-induced misalignment of the pancake vortices that
make up a vortex line. Since YBa$_2$Cu$_3$O$_y$ thin films can be strongly 
disordered, these experiments are sample dependent and don't necessarily 
tell the story in clean samples. In fact, recent mutual inductance 
measurements on 
severely underdoped thin films indicate that YBa$_2$Cu$_3$O$_y$ is really   
quasi-2D only near $T_c$ \cite{Zuev:05}. The weak field dependence of the 
Josephson plasma resonance in YBa$_2$Cu$_3$O$_{6.50}$ single crystals at 
low $T$ is further evidence that the vortices maintain a 3D character 
in clean underdoped samples \cite{Dulic:01}. As for        
underdoped La$_{2-x}$Sr$_x$CuO$_4$, it has been argued from a combined
$\mu$SR and small-angle neutron scattering study of 
La$_{1.9}$Sr$_{0.1}$CuO$_4$ \cite{Divakar:04} that the vortex lines remain 
fairly rigid at low $T$.

\begin{figure}
\includegraphics[width=11.0cm]{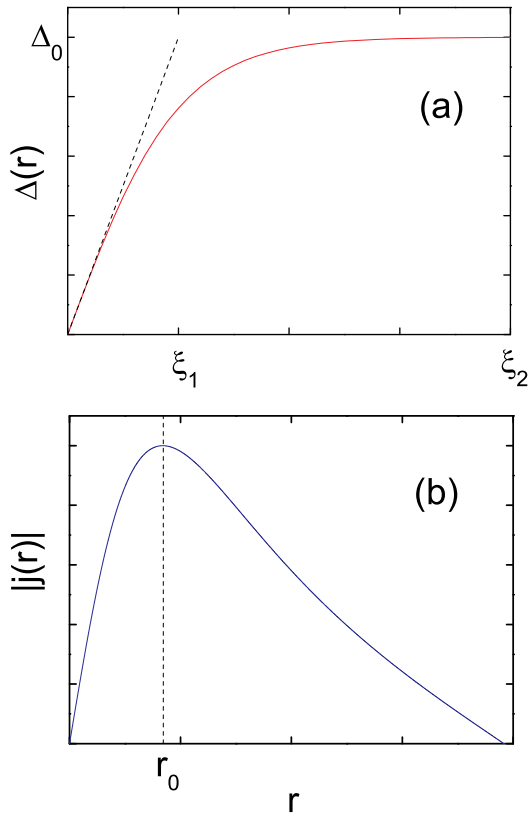}
\caption{Spatial variation of (a) the pair potential $\Delta(r)$ and 
(b) the absolute value of the supercurrent density $j(r)$ for an 
isolated vortex, where $r$ is the radial distance from the vortex centre.}
\label{fig6}
\end{figure}

\section{Vortex core size}
Superconductivity is strongly suppressed in the vortex core. A loose
definition of the vortex core size is that it is a region of radius $r \sim \! \xi$,
where $\xi$ is the characteristic length scale for 
spatial variations of the superconducting order parameter $\psi$---{\it i.e.}
the GL coherence length. However, this definition is not particularly satisfactory,
especially since GL breaks down at low temperatures. Two alternative definitions of 
the vortex core size have emerged from numerical solutions of the Bogoliubov-de Gennes 
and quasiclassical Eilenberger equations.
The first comes from the initial slope of the pair potential $\Delta(r)$ near
the centre of the vortex core. Assuming $\Delta(r) \! \propto \! r$ close
to the core centre, the core radius may be defined as   
\begin{equation}
\xi_1 = \Delta_0 / \lim_{r \rightarrow 0}\frac{\Delta(r)}{r} \, ,
\label{eq:xi1}
\end{equation}
where $\Delta_0 \! \sim \! 1/\xi_0$ is the bulk superconducting energy gap at zero
temperature, and $\xi_0$ is the BCS coherence length.
As shown in figure~\ref{fig6}a, $r \! = \! \xi_1$ is the radius at which
the linear slope extrapolates to $\Delta(r) \! = \! \Delta_0$.
The absolute value of the spatial supercurrent density profile $|j(r)|$ 
provides a second way of defining the core size. As shown in figure~\ref{fig6}b,
$|j(r)|$ exhibits a maximum at a distance $r \! = \! r_0$ from
the core centre.        

\subsection{Magnetic field dependence}

Figure~\ref{fig7} shows the magnetic field dependence of the vortex core size 
determined by $\mu$SR in the $\kappa \! \gg \! 1$ superconductors
V$_3$Si, NbSe$_2$ and LuNi$_2$B$_2$C. Similar results have been reported for
CeRu$_2$ \cite{Kadono:01}, YNi$_2$B$_2$C \cite{Ohishi:02} and 
Nb$_3$Sn \cite{Kadono:06}.
In agreement with calculations from the quasiclassical Eilenberger equations
\cite{Ichioka:99a,Ichioka:99b}, $r_0$ and $\xi_{ab}$ exhibit qualitatively
similar dependences on magnetic field.
As explained in Ref.~\cite{Sonier:04c}, the magnetic field dependence of
$r_0$ is partially due to the `vortex squeezing effect'---which refers to
the increasing overlap of the $j({\bf r})$ profiles of individual vortex 
lines with increasing $H$ (see figure~2 of Ref.~\cite{Sonier:04c}).

It is important to stress that the field dependence of the parameter 
$\xi_{ab}$ that comes from fits of $\mu$SR measurements using 
the models described in section~4.1, does not mean
that the superconducting coherence length is changing with field. Rather, it simply
means that the cutoff factor in these models significantly varies
due to changes in the slope of $\Delta(r)$ in the vortex core region.
For this reason the behaviour of $\xi_{ab}$ is expected to follow that of
$\xi_1$ of equation~(\ref{eq:xi1}), and hence is considered a measure of
the {\it vortex core size}.
In general, the maximum value of the cutoff parameter $\xi_{ab}$ measured by 
$\mu$SR corresponds to the GL coherence length calculated from $H_{c2}$ 
({\it i.e.} $\xi_{\rm GL} \! = \! (\Phi_0/2 \pi H_{c2})^{1/2}$). At low fields, where
the vortices are weakly interacting, the fitted value of $\xi_{ab}$ more or less
agrees with that expected from $H_{c2}$ (see figure~\ref{fig7}).

\begin{figure*}
\includegraphics[width=18.0cm]{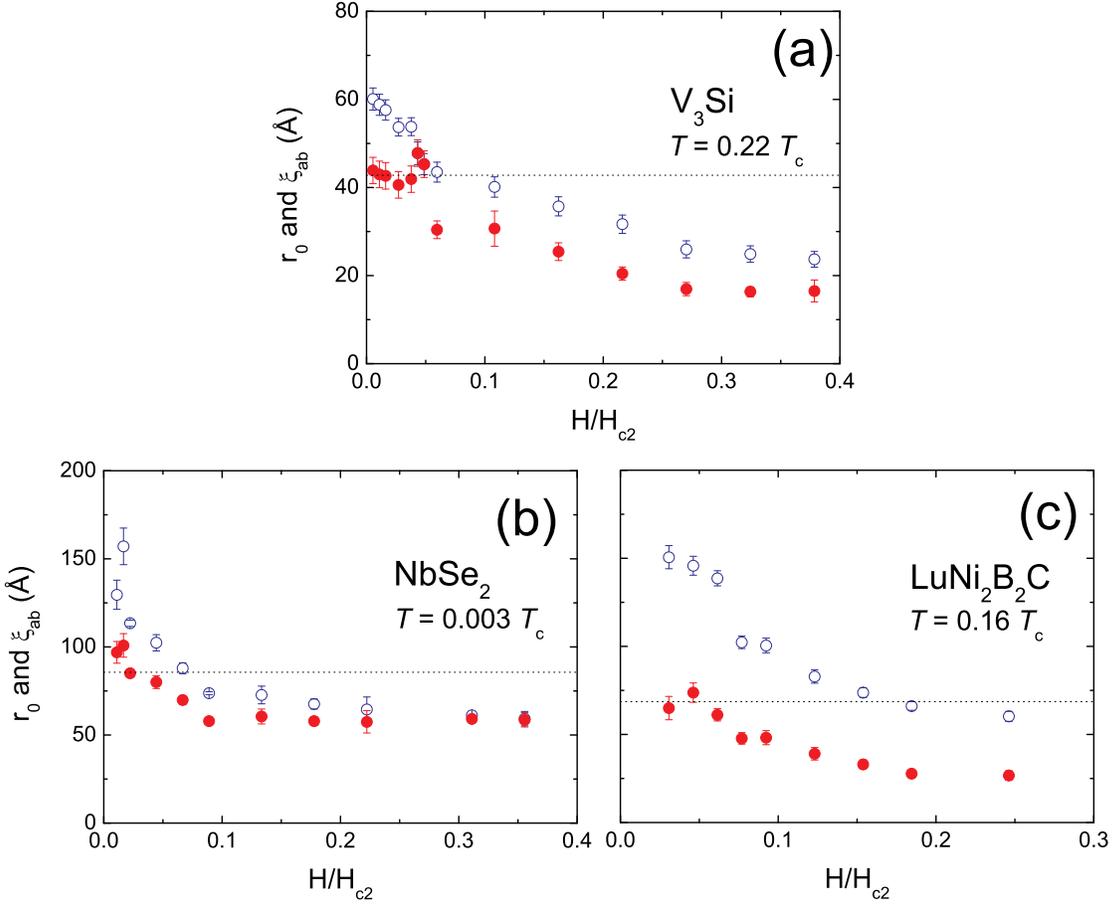}
\caption{Magnetic field dependence of $r_0$ (open circles)
and $\xi_{ab}$ (solid circles) in 
equation~(\ref{eq:Yaouanc}) from $\mu$SR measurements on single
crystals of (a) V$_3$Si \cite{Sonier:04a}, 
(b) NbSe$_2$ \cite{Callaghan:05}, and (c) LuNi$_2$B$_2$C \cite{Price:02}.
All measurements were done with the applied field parallel to the
$\hat{c}$-axis.
The dotted horizontal line in each panel indicates the value of 
the GL coherence length $\xi_{\rm GL}$ calculated from $H_{c2}(T)$.} 
\label{fig7}
\end{figure*}   

\begin{figure}
\includegraphics[width=10.0cm]{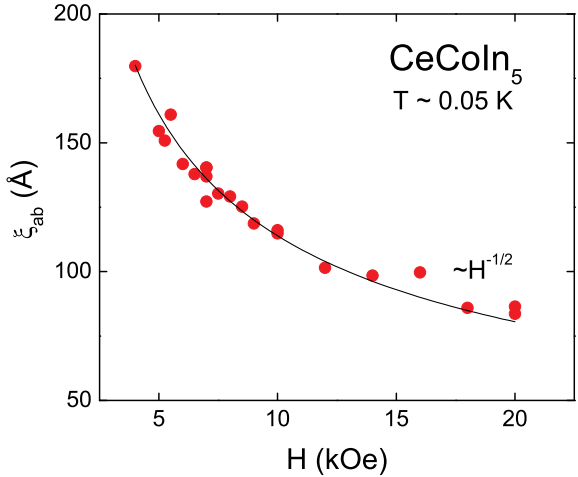}
\caption{Magnetic field dependence of the Ginzburg-Landau coherence
length inferred from the observation of a field-independent FLL form 
factor in small-angle neutron scattering measurements on 
CeCoIn$_5$ \cite{DeBeer:06}.} 
\label{fig8}
\end{figure}   

The observed shrinkage of the vortex cores at higher magnetic fields qualitatively
agrees with theoretical calculations for high-$\kappa$ superconductors
done in the framework of the quasiclassical
Usadel \cite{Sonier:97a,Golubov:94} and Eilenberger equations \cite{Ichioka:99a,Ichioka:99b}.
These theoretical works show that changes in the spatial variation of $j(r)$ (and hence $B(r)$) 
and $\Delta(r)$ with increasing magnetic field are directly related to changes in the electronic 
structure of the vortex cores that occur with increased vortex-vortex interactions.
Physically, the situation is analogous to the electronic band structure
that forms when atomic orbitals are combined in a crystal lattice.  
When atoms come together to form a lattice, the wave functions of the 
electrons anchored to individual atoms overlap the wave functions of the electrons of
neighbouring atoms. This results in the formation of energy bands, which in the case
of a metal, leads to delocalized electrons that are capable of carrying electric 
and thermal currents. In a similar way, as vortices are brought closer
together by increasing the applied magnetic field, there is an increased overlap 
of the wave functions of the single-particle excitations (quasiparticles) 
bound to an individual vortex  \cite{Caroli:64}
with the quasiparticle states of neighbouring vortices 
\cite{Ichioka:99a,Ichioka:99b,Pottinger:93,Tesanovic:94}, resulting in the
intervortex transfer of quasiparticles.
Since the higher-energy bound quasiparticle core states are more spatially
extended, they delocalize first. The reduction of the vortex core size
with increasing magnetic field is thus understood as being due to the
sequential delocalization of quasiparticle core states, starting with
the highest-energy bound state, and terminating with the complete
delocalization of the lowest-energy bound state.

While this picture is strongly supported by a comparison of $\mu$SR
measurements of the vortex core size with the data from other techniques
that probe electronic excitations, Kogan and Zhelezina \cite{Kogan:04} 
have proposed an alternative
theoretical explanation for the field dependence of the core size that is based
on weak-coupling BCS theory. Their model predicts that in clean high-$\kappa$
superconductors, the intrinsic superconducting coherence length is not
independent of magnetic field, but actually decreases with increasing field.
Recently, DeBeer-Schmitt {\it et al.} \cite{DeBeer:06}
measured an unusual field-independent FLL form factor in CeCoIn$_5$
by small-angle neutron scattering, suggestive of a field-dependent
coherence length (see figure~\ref{fig8}). Even so, there may be another
explanation for their data that warrants further experimental investigation of
the FLL form factor in other materials.     
A seemingly direct test of the Kogan-Zhelezina theory would be to 
study the magnetic field dependence of the core size in impurity-doped 
superconductors. Unfortunately, the 
addition of impurities is likely to interfere with the intervortex transfer
of quasiparticles and add disorder to the FLL. Thus the interpretation
of such experiments would probably be ambiguous.
 
\subsection{Specific heat}
Specific heat measurements are sensitive to both localized and delocalized
quasiparticle excitations. In the vortex state the specific heat is usually
described by
\begin{equation}
c(T, B) = \gamma(T, B)T + \alpha T^2 + \beta T^3 \, ,
\label{eq:heatH}
\end{equation}
where the last term $\beta T^3$ is the phonon contribution.
The first term is the electronic contribution to the zero-energy quasiparticle
density of states (DOS) $N(E \! = \! 0)$, {\it i.e.} the spatial average of the
local DOS $N(E \! = \! 0, {\bf r})$ at the Fermi level $E \! = \! 0$. 
The `Sommerfeld coefficient' is given as
\begin{equation}
\gamma(B) \! \equiv \! {\rm lim}_{T \rightarrow 0} \, c(T, B)/T \propto N(0) \, .
\end{equation}
The second term in equation (\ref{eq:heatH}) is not always clearly present in 
experimental data, and only recently has theoretical progress in understanding
its origin been made. Nakai {\it et al} \cite{Nakai:06} have shown there
are two electronic contributions to the $\alpha T^2$ term. The first comes from 
low-energy states ($E \! > \! 0$) that appear near the Fermi energy at 
$B \! \neq \! 0$. The second contribution comes from the `Kramer-Pesch effect',
which is a shrinking of the vortex cores with decreasing temperature that is
associated with the thermal depopulation of the bound quasiparticle core states.
As discussed elsewhere \cite{Sonier:00,Sonier:04c}, 
the Kramer-Pesch effect has been measured by $\mu$SR in a variety of
superconductors.  

In an isotropic-gapped superconductor, the low-lying excitations are generally 
considered to be confined to the vortex cores \cite{Caroli:64}. In this case
$\gamma(B)$ is proportional to the product of the area of a vortex core 
$\pi \xi^2$ and the density of vortices \cite{Fetter:69}, so that
\begin{equation}
\gamma(B) \! \propto \! \pi \xi^2 B \, . 
\end{equation}
Thus, if the size of the vortex cores is independent of magnetic field,
then $\gamma(B) \! \propto \! B$. 
In highly anisotropic-gapped 
superconductors, including those with line or point nodes at the Fermi
surface, the low-lying excitations near the gap minima become the dominate
contribution to $\gamma(B)$. Volovik showed that for the case of a $d$-wave
superconductor, the superfluid flow around a vortex lowers the energy of 
quasiparticles delocalized near the gap nodes, leading to a finite DOS at 
the Fermi energy and a corresponding nonlinear field dependence for $\gamma(B)$ 
\cite{Volovik:93}. For the case of a highly anisotropic energy gap without nodes, 
such behavior is expected only if the energy shift exceeds the gap minimum.
Until recently, this is more or less how specific heat measurements on type-II 
superconductors as a function of magnetic field were interpreted. 
$\gamma(B) \! \propto \! B$ behaviour was indicative of conventional $s$-wave
superconductivity, and a sublinear dependence of $\gamma(B)$ on field, usually
$\gamma \! \propto \sqrt{B}$, meant unconventional superconductivity.

In 1999 it became clear that the delocalization of quasiparticles
brought about by vortex-vortex interactions shows up in the field
dependence of the specific heat of $s$-wave superconductors---often
mimicking the behaviour expected for an unconventional pairing-state symmetry.
Theoretically it was shown from solutions of the quasiclassical Eilenberger 
equations at $T \! = \! 0.5T_c$ that the zero-energy DOS per vortex $N(0)/B$ 
depends on magnetic field \cite{Ichioka:99a}. In the same work this
unexpected behaviour was shown to be due to a shrinking of the vortex cores
with increasing field, associated with an increased overlap of the 
zero-energy DOS of one vortex with that of its neighbours. 
This idea was experimentally verified
in the same year by relating the field dependence of the vortex
core size measured in NbSe$_2$ by $\mu$SR \cite{Sonier:97a} 
to the field dependence of the specific heat of this material 
\cite{Sonier:99a}. 

More recent calculations in the framework of the
quasiclassical Eilenberger theory  \cite{Nakai:04} show that for an
$s$-wave superconductor at $T \! = \! 0.1T_c$, the isolated vortex behaviour 
$\gamma(B) \! \propto \! B$ is in fact realized at low temperatures
up to a crossover field $B^*$. Above $B^*$, $\gamma(B)$ has a sublinear 
dependence on field due to the overlap of the low-energy 
quasiparticle core states from neighbouring vortices.
Even so, it is rare to find superconducting materials that exhibit
$\gamma(B) \! \propto \! B$ behaviour at low magnetic fields, 
because any anisotropy of the superconducting energy gap or Fermi surface
reduces the value of $B^*$. This also means that the vortex core size
measured by $\mu$SR will most often be dependent on field.
Recently, this was shown to be the case even at low field
in the marginal type-II superconductor V \cite{Laulajainen:06}. 
The field dependence of the core size in pure V cannot be explained by 
the theory of Ref.~\cite{Kogan:04}, which does not apply to low-$\kappa$ 
superconductors. On the other hand, an explanation in terms of
delocalized core states is compatible with the sublinear dependence
of $\gamma(B)$ on $B$ found immediately above $H_{c1}$ in the 
low-$\kappa$ type-II superconductor Nb \cite{Sonier:06a}. 
       
\subsection{Thermal conductivity}

In contrast to the electronic contribution to the specific heat in the mixed
state, the electronic contribution to heat transport 
comes entirely from extended or delocalized quasiparticles. 
In superconductors with extreme gap anisotropy, such as LuNi$_2$B$_2$C 
\cite{Boaknin:01} or YBa$_2$Cu$_3$O$_{y}$ \cite{Chiao:99}, 
the dominant contribution to the field dependence
of the thermal conductivity at low $T$ is the Doppler shift of the
quasiparticle spectrum outside the vortex cores
due to the circulating supercurrents \cite{Kubert:98,Vekhter:99}.
However, for an $s$-wave superconductor, the field dependence of
the electronic thermal conductivity $\kappa_e$ obtained by extrapolating 
measurements to $T \! \rightarrow \! 0$~K is a direct measure of the
delocalization of quasiparticle states bound to the vortex cores.

In recent years, $\mu$SR measurements on $s$-wave superconductors
have established a direct correlation between the field dependence of
the vortex core size and the delocalization of quasiparticle core
states as measured by thermal conductivity. By imposing the following simple 
model, excellent agreement between these two very different kinds of
measurements has been demonstrated \cite{Callaghan:05}:\\

\noindent The electronic thermal conductivity $\kappa_e$ is proportional to the
zero-energy density of {\it delocalized} states $N_{\rm deloc}(0)$,
and hence grows as the zero-energy density of bound cores states $N(0)$ 
decreases. Since $N(0)$ is proportional to $\pi \xi^2 B$, then
\begin{equation}
\kappa_e \propto N_{\rm deloc} \propto (\pi \xi(B_{c1})^2 - \pi \xi(B)^2)B \, ,
\label{eq:CoreAreaReduced}
\end{equation} 
where $\pi \xi(B_{c1})^2$ and $\pi \xi(B_{c1})^2$ are the areas of the 
vortex core at $B \! = \! B_{c1}$ and $B \! > \! B_{c1}$, respectively.

\subsubsection{V$_3$Si: a good standard.}
In 2004, a $\mu$SR study of $n(B)$ in the mixed state of the high-$\kappa$ 
$s$-wave superconductor V$_3$Si was reported \cite{Sonier:04a}. The results
of this experiment provide a good standard for comparison of $\mu$SR
measurements on more complicated systems. The reason is that at low field,
the electronic states remain fairly well localized within the vortex cores.
Consequently, $\kappa_e(T \! \rightarrow \! 0, H)$ increases slowly
with increasing magnetic field \cite{Boaknin:03}, $c(T \! \rightarrow \! 0, H)$
is dominated by localized electronic states at low $H$, and
the vortex core size determined by $\mu$SR is essentially independent
of $H$ up to $H \! \approx \! 7.5$~kOe. As shown in figure~\ref{fig9}a,
equation~(\ref{eq:CoreAreaReduced}) accurately describes the
thermal conductivity data using the $\mu$SR values of the
vortex core size $\xi_{ab}$.

\begin{figure*}
\includegraphics[width=18.0cm]{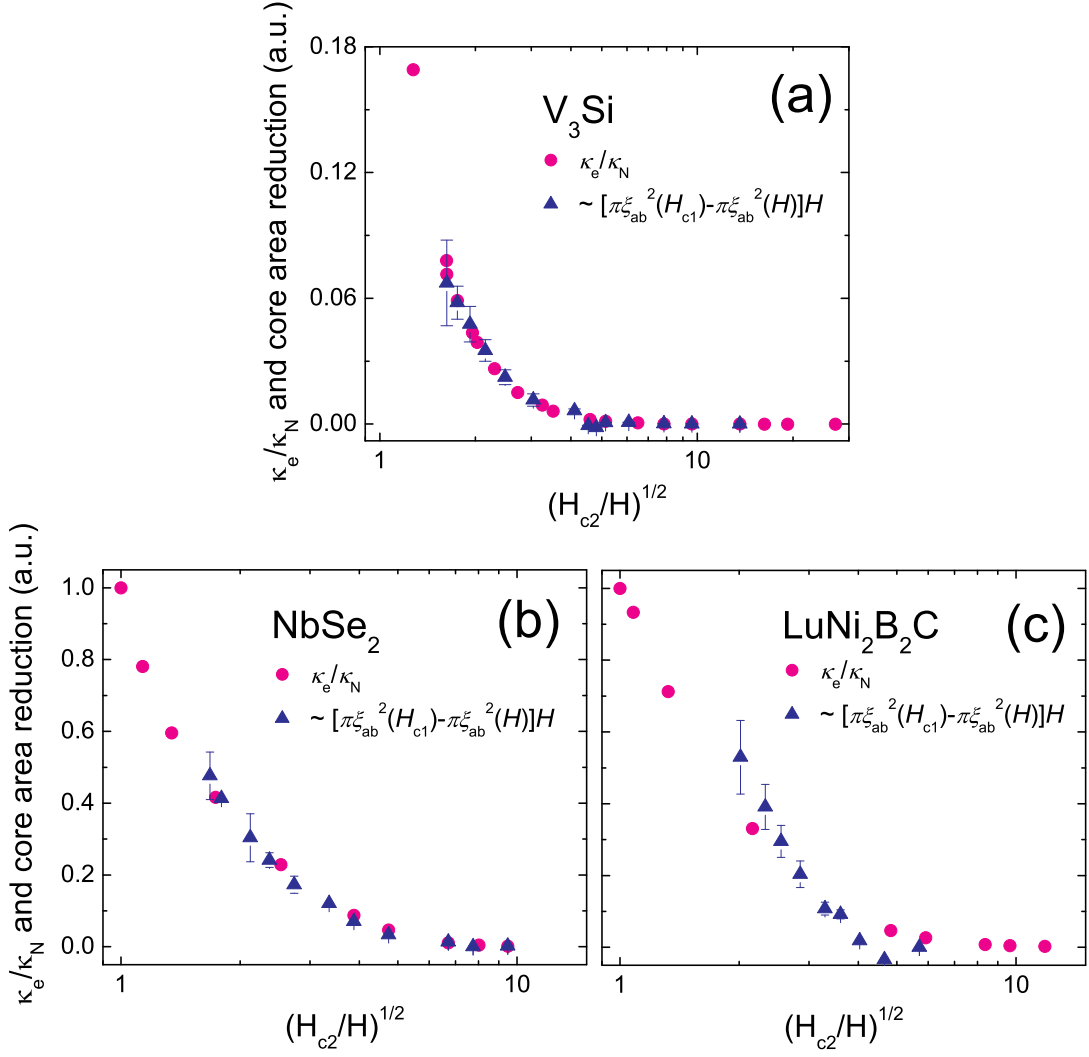}
\caption{(a) $\mu$SR \cite{Sonier:04a} and electronic thermal conductivity 
\cite{Boaknin:03} data for V$_3$Si. The electronic thermal conductivity 
data (solid circles) is normalized to the value $\kappa_{\rm N}$
at $H_{c2}$. The $\mu$SR data (solid triangles) is plotted as
equation~(\ref{eq:CoreAreaReduced}).
(b) Equivalent $\mu$SR \cite{Callaghan:05} and electronic thermal conductivity 
\cite{Boaknin:03} data for NbSe$_2$. 
(c) Equivalent $\mu$SR \cite{Price:02} and electronic thermal conductivity 
\cite{Boaknin:01} data for LuNi$_2$B$_2$C.}
\label{fig9}
\end{figure*} 

\subsubsection{V and Nb: marginal type-II superconductors.}
As mentioned earlier, the vortex core size in the low-$\kappa$ superconductor 
V was recently measured by $\mu$SR \cite{Laulajainen:06}. With the magnetic 
field applied parallel to the $\langle 111 \rangle$ direction of a 
V single crystal, the vortex core size was found to decrease immediately above 
the lower critical field
$H_{c1} \! \approx \! 1$~kOe. The high degree of delocalization of quasiparticle 
core states necessary to cause the observed shrinking of the vortex core size at 
such low field is a consequence of the large superconducting coherence
length ($\sim 300$~\AA) and 10~\% to 20~\% gap anisotropy. 

While there are
no reported measurements of $\kappa_e(T \! \rightarrow \! 0, H)$ for V,
there are for the related elemental superconductor Nb \cite{Lowell:70}.
The electronic thermal conductivity of Nb reported in Ref.~\cite{Lowell:70}
behaves much like that of V$_3$Si, exhibiting a weak exponential increase
above $H_{c1}$. This is clearly at odds with the $\mu$SR results for V.
However, it is also at odds with the specific heat measurements of Nb 
reported in Ref.~\cite{Sonier:06a}, which indicate an appreciable overlap
of the quasiparticle core states of neighbouring vortices immediately above
$H_{c1}$. As argued in Ref.~\cite{Sonier:06a}, this discrepancy is
due to the way in which these experiments were done. When the electronic
specific heat of Nb was measured under zero-field cooled conditions, no 
appreciable contribution from localized or delocalized quasiparticle 
states was observed at fields near $H_{c1}$. This is consistent with the thermal 
conductivity data of Ref.~\cite{Lowell:70} measured in monotonically 
increasing and decreasing magnetic field. On the other hand, under
the same field-cooled conditions implemented in the $\mu$SR study of V,
the specific heat measurements are consistent with delocalized quasiparticle
states immediately above $H_{c1}$. These experiments show that
the intervortex transfer of quasiparticles is disrupted when the FLL is
highly disordered. While a well-ordered FLL forms in the sample under
field-cooled conditions, flux entry into the sample is impeded by the 
Bean-Livingston barrier under zero-field cooled or monotonically increasing
field conditions. The field dependence of $\kappa_e(T \! \rightarrow \! 0, H)$ 
has yet to be measured in Nb or V under field-cooled conditions. 

\subsubsection{NbSe$_2$: multi-band superconductivity.}
For a long time the mixed state of NbSe$_2$ was considered an archetype of 
the FLL in conventional type-II superconductors. However, over the past 6 years 
a number of experiments on NbSe$_2$ have provided strong evidence for distinct 
energy gaps on different Fermi sheets \cite{Boaknin:03,Yokoya:01,Rodrigo:04}.
While there is some debate over whether the smaller energy gap resides
on the Se or Nb Fermi sheets \cite{Carrington:06}, the experimental signatures
of multi-band superconductivity are similar to those of MgB$_2$.

Extending theoretical work on MgB$_2$ 
\cite{Nakai:02b,Koshelev:03,Dahm:04,Ichioka:04} to the case of NbSe$_2$, 
suggests that the nonlinearity of $\gamma(B)$ observed at low $B$ can be 
attributed to the occurrence of a smaller energy gap $\Delta_0$ on one of 
the Fermi surfaces. In these theories the electronic states of a vortex
are dependent on the combined contributions from two different bands, 
characterized by large and small superconducting energy gaps and 
different Fermi surface anisotropies.
At low magnetic field the dominant contribution to the low-energy
quasiparticle core states comes from the small-gap band. Because of the
small size of the energy gap, these core states are loosely bound to the 
vortex and easily delocalize via overlap with the low-energy core states of
neighbouring vortices. Consequently, the vortex core is large at low
field, and shrinks with increasing magnetic field. However, in contrast
to a single-band superconductor, low-energy quasiparticle states
continue to be localized around the the vortex cores at high fields
due to the contribution from the large-gap band. In particular,
the large-gap band results in electronic states that are more
confined to the vortex core. While the electronic vortex core states of
the individual bands can be probed separately by STS \cite{Eskildsen:02}, 
this does not
imply that there are two different kinds of vortices separated in real space.
Both bands contribute to a single vortex, but the magnetic
field dependence of each contribution is different. 
Moreover, the energy gaps of the two bands are related in the sense 
that only a single superconducting transition occurs.    

\begin{figure}
\includegraphics[width=10.0cm]{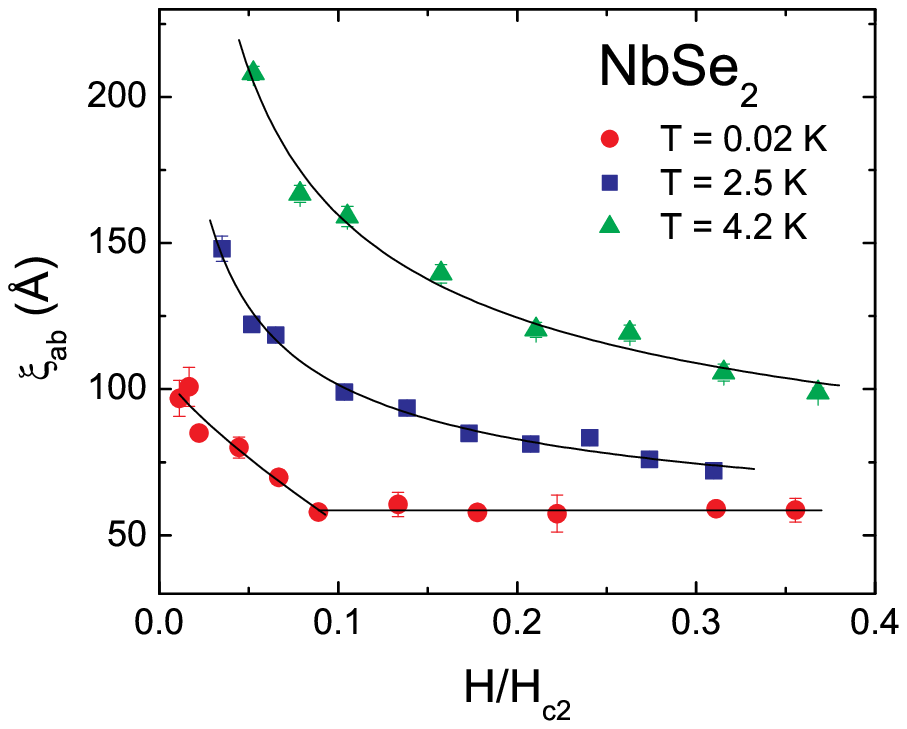}
\caption{Magnetic field dependence of the cutoff parameter 
$\xi_{ab}$ from $\mu$SR measurements on single
crystals of NbSe$_2$ at different temperatures 
\cite{Sonier:97a,Callaghan:05}.} 
\label{fig10}
\end{figure}   
 
In a $\mu$SR study of NbSe$_2$ at $T \! = \! 0.02$~K, in which thermal excitations
of the bound quasiparticle core states are largely frozen out, 
Callaghan {\it et al.} \cite{Callaghan:05} explicitly detected large vortex cores 
at low magnetic fields (see figure~\ref{fig10}). With increasing $H$ the core size 
rapidly shrinks. This behaviour is attributed to the ease at which the loosely 
bound core states delocalize via their overlap with the core states of 
neighbouring vortices. As shown in figure~\ref{fig10}, the core size saturates 
above $H \! \approx \! 0.1 H_{c2}$. 
At these fields the experiment probes the smaller 
vortices associated with the large-gap band. The saturation of the core size 
reflects the reluctance of the tightly bound core states of these smaller 
vortices to delocalize. This interpretation of the $\mu$SR measurements 
is in accord with what is known about the vortices in MgB$_2$. 
At fields far below $H_{c2}$, Eskildsen {\it et al.} \cite{Eskildsen:02} detected
vortices in MgB$_2$ by STS that have a large core size relative to estimates 
from $H_{c2}$ and do not have localized core states. This is consistent with
the full field dependence of the core size inferred from specific heat
measurements \cite{Klein:06}. The unavailability of single 
crystals of sufficient size has so far prevented a study 
of $\xi_{ab}(H)$ in MgB$_2$ by $\mu$SR.      

The two-gap interpretation of the field dependence of the core size in NbSe$_2$ 
is also supported by a remarkable agreement with the 
field dependence of $\kappa_e(T \rightarrow 0, H)$ measured independently 
in Ref.~\cite{Boaknin:03} (see figure~\ref{fig9}b). In contrast to V$_3$Si,
$\kappa_e(T \rightarrow 0, H)$ rises rapidly with increasing field 
just above $H_{c1}$, indicating the presence of highly delocalized quasiparticle 
states. This is the same behaviour that is observed for the electronic thermal 
conductivity of MgB$_2$ \cite{Sologubenko:02} and the heavy-fermion
compound PrOs$_4$Sb$_{12}$ \cite{Seyfarth:05,Seyfarth:06}, and is understood
to arise from two or more superconducting energy gaps. As is the case for
MgB$_2$, intermediate size single crystals of PrOs$_4$Sb$_{12}$ are not 
yet available for $\mu$SR studies, and hence the field dependence of the vortex 
core size has not been measured. 

\begin{figure}
\includegraphics[width=10.0cm]{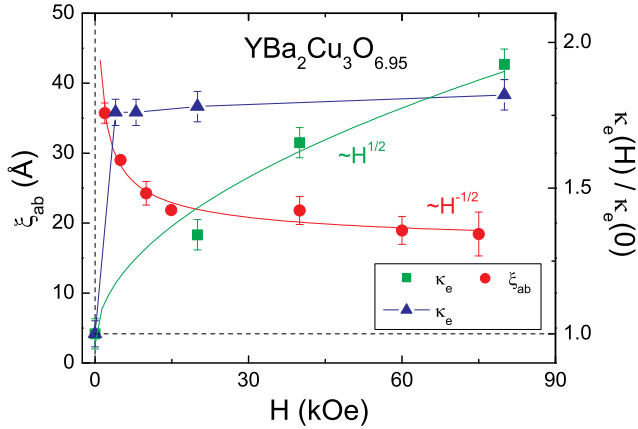}
\caption{The magnetic field dependence of $\xi_{ab}$ in YBa$_2$Cu$_3$O$_{6.95}$
determined by $\mu$SR \cite{Sonier:99b} and the normalized
electronic thermal conductivity $\kappa_e(H)/\kappa_e(0)$ measured in
YBa$_2$Cu$_3$O$_{6.9}$ by Chiao {\it et al.} \cite{Chiao:99} and in
YBa$_2$Cu$_3$O$_7$ by Hill {\it et al.} \cite{Hill:04}, respectively.
All data are extrapolations to $T \! \rightarrow \! 0$~K.
Note, the field dependence of $\xi_{ab}$ shown here is
a more accurate analysis of the measurements of Ref.~\cite{Sonier:99b}. 
In particular, the modified Bessel function $K_1(u)$ in
equation~(\ref{eq:Yaouanc}) was calculated
numerically, whereas the analysis of Ref.~\cite{Sonier:99b}  
assumed the following analytical approximation:
$K_1(u) \! = \! (\pi/2u)^{1/2} \exp(-u)$, valid for $u \! \gg \! 1$.}
\label{fig11}
\end{figure} 

\subsubsection{YBa$_2$Cu$_3$O$_y$: highly anisotropic and multiple gaps?}
As shown in figure~\ref{fig11}, the vortex-core size in   
the high-$T_c$ superconductor YBa$_2$Cu$_3$O$_{6.95}$ \cite{Sonier:99b} shrinks 
with increasing magnetic field, such that 
$\xi_{ab}(H)-\xi_{ab}(0) \! \propto \! H^{-1/2}$. At first
glance this behaviour resembles the previous examples, where the shrinking
of the vortex cores has been associated with the delocalization of quasiparticle
core states. However, in YBa$_2$Cu$_3$O$_y$ the low-energy 
quasiparticle core states are expected to be
extended along the nodal directions of the $d_{x^2-y^2}$-wave 
gap function \cite{Ichioka:96,Ichioka:99b}, and hence are already
delocalized at low field. Furthermore, the value of $\xi_{ab}$ deduced by
$\mu$SR at low $H$ seems rather large. As discussed in Ref.~\cite{Sonier:04c}, 
the large vortex-core size at low $H$ appears to be unique among the cuprates and 
likely results from {\it proximity-induced} superconductivity on the CuO chains.
Calculations by Atkinson \cite{Atkinson:07} show that large
two-fold symmetric vortices result from coupling of the CuO chain
and CuO$_2$ plane layers (see figure~\ref{fig12}).   
The vortex core size in YBa$_2$Cu$_3$O$_{6.95}$ saturates near 
$H \! = \! 40$~kOe, at which the FLL begins a gradual hexagonal-to-square
transition \cite{Brown:04}. The $d_{x^2-y^2}$-wave superconducting order 
parameter is predicted to produce fourfold symmetry in the spatial field 
profile $B({\bf r})$ around the vortex cores \cite{Ichioka:96}. Thus 
the fourfold symmetry of the FLL at high fields likely results from 
strong overlap of the intrinsic $d_{x^2-y^2}$-wave vortices.            

\begin{figure*}
\includegraphics[width=16.0cm]{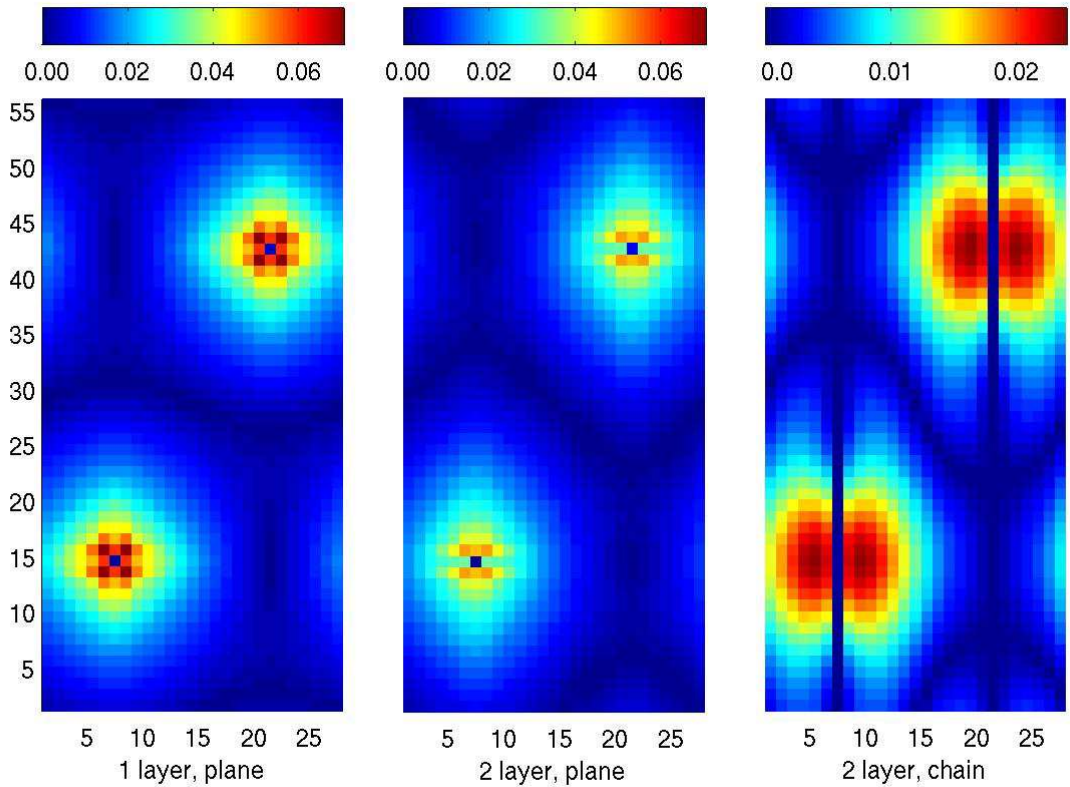}
\caption{Contour plot of the spatial dependence of the magnitude of the
supercurrent density $j({\bf r})$ for vortices in a single-layer
$d_{x^2-y^2}$-wave superconductor (left) calculated from self-consistent 
solutions of the Bogoliubov de Gennes equations \cite{Atkinson:07}.
Also shown are the results of similar calculations for a
2 layer superconductor, consisting of a {\it chain} layer coupled to a
single $d_{x^2-y^2}$-wave superconducting {\it plane}. The middle panel shows 
what the vortices look like on the plane layer, and the right panel shows
what they look like on the chain layer.} 
\label{fig12}
\end{figure*}

A comparison of the $\mu$SR data to electronic thermal conductivity measurements
by Chiao {\it et al.} \cite{Chiao:99} seems to imply that there are bound core states
which delocalize with increasing magnetic field (see figure~\ref{fig11}).
However, the behaviour of $\kappa_e(H)$ is more likely due to a field-induced
Doppler shift of the quasiparticle spectrum outside the vortex cores 
\cite{Kubert:98,Vekhter:99}. More recent measurements on
ultraclean single crystals of YBa$_2$Cu$_3$O$_7$ by Hill {\it et al.} \cite{Hill:04}
show a saturation of $\kappa_e(H)$ at low $H$, argued to indicate an exact
cancellation of the contributions from the Doppler shift and scattering
of the quasiparticles by the vortices. Thus, consistent with a highly 
anisotropic gap, there is no established 
correlation between the size of the vortex cores 
and electronic thermal conductivity measurements on very clean samples.      
 
\subsubsection{$R$Ni$_2$B$_2$C: highly anisotropic or multiple gaps?}
For some time now, experimental evidence has mounted that the superconducting 
energy gap in rare-earth nickel borocarbides $R$Ni$_2$B$_2$C 
($R \! \equiv$ Dy, Ho, Er, Tm, Lu and Y) is highly anisotropic, with
line or point nodes. For example, thermal conductivity measurements on
LuNi$_2$B$_2$C reveal the existence of highly delocalized quasiparticles 
above $H_{c1}$ \cite{Boaknin:01}. However, there is also evidence suggesting 
that these borocarbide materials exhibit multi-band superconductivity
\cite{Shulga:98,Doh:99,Mukhop:05,Huang:06}. Figure~\ref{fig9}c shows 
that unlike YBa$_2$Cu$_3$O$_y$, for LuNi$_2$B$_2$C there is good agreement between 
$\mu$SR measurements of the vortex core size and the field dependence of the 
electronic thermal conductivity. This indicates that there are quasiparticle
vortex-core states at low temperatures that delocalize 
with increasing magnetic field.
Since the anisotropy of the superconducting energy gap manifests itself in the 
electronic structure of the vortices, the superconducting gap of LuNi$_2$B$_2$C
must be less anisotropic than in YBa$_2$Cu$_3$O$_y$. 
Moreover, measurements done to date are not inconsistent with multiple 
superconducting gaps. The $\mu$SR measurements on LuNi$_2$B$_2$C 
by Price {\it et al.} \cite{Price:02} only extend down to $T \! = \! 2.5$~K. 
As in the case of NbSe$_2$, a clear
signature of multiple-gap superconductivity might be seen in the field
dependence of the vortex-core size at lower temperatures, where thermal 
excitations of the quasiparticle core states are largely frozen out.
Interestingly, $\mu$SR measurements of $\xi_{ab}(H)$ in YNi$_2$B$_3$C at 
$T \! = \! 3$~K by Ohishi {\it et al.} \cite{Ohishi:02} display a clear saturation
above $H \! \approx \! 5$~kOe indicative of a second superconducting gap.     
 
\subsection{Magnetic field dependence of $\lambda_{ab}$}

It is important to stress that the field dependence of the fit
parameter $\lambda_{ab}$ 
generally does not imply that the superfluid density depends on field.
Instead it is an indication that the theoretical model for $n(B)$ does
not contain all of the relevant physics. On the other hand,
a ``true'' measure of the {\it magnetic penetration depth} in a 
$\mu$SR experiment has been demonstrated to be given by the
$H \! \rightarrow \! 0$ extrapolated value of $\lambda_{ab}$
\cite{Callaghan:05,Laulajainen:06}. In other words, when an inadequate
model for $n(B)$ is used to fit the TF-$\mu$SR signal, the difference
between $\lambda_{ab}$ and the magnetic penetration depth grows with
increasing magnetic field. This has been nicely demonstrated in several
theoretical works \cite{Amin:98,Amin:00,Laiho:06,Laiho:07}.   

The {\it effective} length scale $\lambda_{ab}$ measured by $\mu$SR in
the vortex state of single-crystal superconductors usually exhibits a strong dependence
on magnetic field at low temperatures. 
This was first observed in YBa$_2$Cu$_3$O$_{6.95}$ single
crystals \cite{Sonier:97c}, where it was attributed to effects associated with the 
high anisotropy of the $d_{x^2-y^2}$-wave superconducting energy gap $\Delta({\bf k})$.
There are two primary effects of strong gap and/or Fermi surface anistropy on 
$\lambda_{ab}$:

\begin{itemize}
\item A nonlinear supercurrent response to the applied field $H$ results from
a quasi-classical `Doppler' shift of the quasiparticle energy spectrum
by the flow of superfluid around a vortex \cite{Volovik:93}. When the
Doppler shift exceeds the energy gap, Cooper pairs are broken,
and $\lambda$ increases. Nonlinear effects are particularly strong in a  
$d_{x^2-y^2}$-wave superconductor, where even at low $T$ and weak $H$
pair breaking occurs from the excitation of quasiparticles at
the gap nodes. A sizeable nonlinear response can also occur in $s$-wave 
superconductors that have a highly-anisotropic energy gap, a 
smaller energy gap on one of the Fermi sheets ({\it i.e.} a multi-band
superconductor), and/or a highly-anisotropic Fermi surface.                              

\item A nonlocal supercurrent response to the applied field $H$ occurs
in the limit $\lambda \! << \! \xi_0$, meaning that the supercurrent
density ${\bf j}({\bf r})$ depends on the magnetic vector potential
${\bf A}({\bf r})$ within a volume of radius $\sim \xi_0$ around ${\bf r}$.
Since $\xi_0 \! = \! \hbar v_{\rm F} / \pi \Delta$, the value of the
superconducting coherence length is ${\bf k}$-dependent, {\it i.e.}
it is dependent on both the energy gap $\Delta({\bf k})$ and
the Fermi velocity $v_{\rm F}({\bf k})$. In a $d_{x^2-y^2}$-wave 
superconductor, nonlocal effects arise from the divergence of $\xi_0({\bf k})$
at the gap nodes, but they may also occur in isotropic-gapped 
superconductors having strong Fermi surface anisotropy. In either situation,
the nonlocal response vanishes far from the vortex cores, so that only
the spatial distribution of magnetic field in and around the vortex cores 
is influenced. 
\end{itemize}

\begin{figure}
\includegraphics[width=10.0cm]{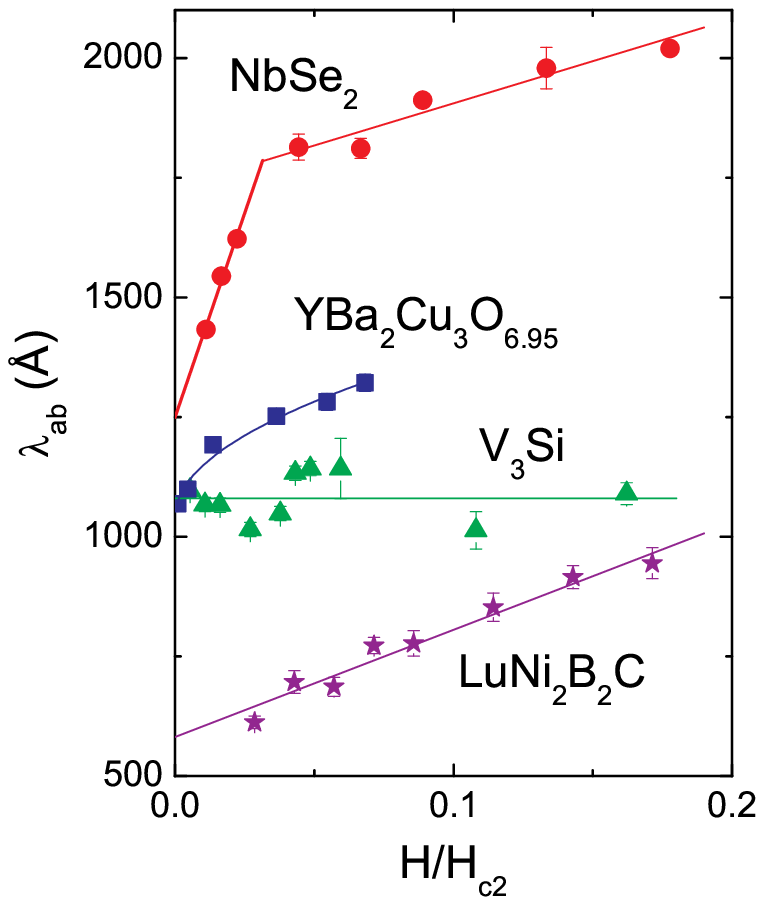}
\caption{Magnetic field dependence of the fitting parameter
$\lambda_{ab}$ plotted versus reduced field for LuNi$_2$B$_2$C \cite{Price:02},
V$_3$Si \cite{Sonier:04a}, NbSe$_2$ \cite{Callaghan:05} and
YBa$_2$Cu$_3$O$_{6.95}$ \cite{Sonier:99a}. The reduced field for 
YBa$_2$Cu$_3$O$_{6.95}$ assumes $H_{c2} \! = \! 1100$~kOe, as determined
in Ref.~\cite{Nakagawa:98}. The results for V$_3$Si and LuNi$_2$B$_2$C 
come from fits to the `Kogan model' \cite{Kogan:97a,Kogan:97b}
that includes nonlocal effects,
whereas the data for NbSe$_2$ and YBa$_2$Cu$_3$O$_{6.95}$ come
from fits assuming an analytical GL model \cite{Yaouanc:97}.}   
\label{fig13}
\end{figure}

In 1999, a high-field TF-$\mu$SR study of YBa$_2$Cu$_3$O$_{6.95}$ \cite{Sonier:99b} 
showed that $\lambda_{ab}(H)$ has a sublinear dependence on $H$, in apparent agreement 
with theoretical predictions for nonlocal and nonlinear effects 
in the vortex state of a $d_{x^2-y^2}$-wave superconductor \cite{Amin:98}. In this
case the dominant contribution to the field dependence of $\lambda_{ab}(H)$ are
nonlocal effects. With increasing $H$, the increased overlap of the 
spatial regions around the vortex cores characterized by a nonlocal response, 
reduces the width of the $\mu$SR line shape.
Since the phenomenological GL and London models for $B({\bf r})$ do not account
for this, the fitted value of $\lambda_{ab}$ exceeds the actual magnetic penetration
depth. 

There have been several $\mu$SR studies on systems that exhibit field-induced
hexagonal-to-square FLL transformations, where the $\mu$SR signals have been
fit assuming a phenomenological London model developed by
Kogan {\it et al.} \cite{Kogan:97a,Kogan:97b}. The Kogan model
includes nonlocal corrections that stem from Fermi surface anisotropy.
As shown in figure~\ref{fig13}, the Kogan model adequately accounts
for nonlocal effects in V$_3$Si at low field, as the fitted
$\lambda_{ab}$ is field-independent. The same model has also been
applied to $\mu$SR measurements on borocarbide superconductors 
\cite{Ohishi:02,Price:02} and Nb$_3$Sn \cite{Kadono:06}. 
However, in these cases $\lambda_{ab}$ still exhibits a field dependence. 
This suggests that there
is a highly anisotropic superconducting gap (or multiple gaps) that
causes the field dependence of $\lambda_{ab}$---which in turn 
indicates that the theoretical model fails to accurately portray the
field distribution of the FLL. 

Kadono \cite{Kadono:04c} has argued that the field dependence of $\lambda_{ab}$
measured by $\mu$SR in any type-II superconductor is indicative of the 
degree of anisotropy of the superconducting order parameter. This
seems to be true, although $\lambda_{ab}(H)$ is partly influenced 
by changes in the field 
decay outside the vortex core arising from an increased overlap of the
quasiparticle core states of neighbouring vortices. Some examples of the
field dependence of $\lambda_{ab}$ 
obtained in $\mu$SR studies are shown in figure~\ref{fig13}.
For the case of V$_3$Si, where the quasiparticles are tightly bound to
the vortex cores at low reduced field $H/H_{c2}$, $\lambda_{ab}$ is
at most weakly dependent on field. The Fermi surface anisotropy responsible
for the square FLL that occurs at higher fields does not affect
$\lambda_{ab}(H)$ at low field. On the other hand, the rapid delocalization
of quasiparticles in NbSe$_2$ with increasing magnetic field appears to
influence $\lambda_{ab}(H)$, as it exhibits a stronger field dependence
than what is observed in highly anisotropic YBa$_2$Cu$_3$O$_{6.95}$.
Recently, Laiho {\it et al.} \cite{Laiho:07} have shown that the field dependences
of $\lambda_{ab}$ in V$_3$Si and NbSe$_2$ vanish if an appropriate 
temperature and field-dependent cutoff function numerically calculated 
from the quasiclassical Eilenberger equations for an $s$-wave
superconductor is used to fit the $\mu$SR
measurements. This is because the low-energy excitations of the vortex 
cores that affect the field distribution are handled by the
quasiclassical Eilenberger theory. While the magnetic penetration 
depth of an isotropic $s$-wave superconductor in the vortex state 
has no appreciable field dependence in the microscopic theory,  
the vortex core size has a strong field dependence due to
the intervortex transfer of quasiparticles \cite{Ichioka:99a,Ichioka:99b}.            
 
\section{Magnetism in High-$T_c$ Superconductors}

At the turn of the 21$^{\rm st}$ century, neutron scattering, NMR 
and $\mu$SR experiments on high-$T_c$ superconductors revealed that 
the application of a magnetic field may induce or enhance 
antiferromagnetic (AF) spin correlations. In many cases, the AF spin 
correlations are most pronounced in and around the vortex cores
where superconductivity is suppressed. These findings suggest 
there is a competing magnetic order that coexists with 
superconductivity. However, the theoretical implications of this 
hinge on some important details about the experiments, such as whether 
the observed magnetism is static (or quasistatic) or dynamic, and 
whether or not static (or quasistatic) magnetism is also
present in zero magnetic field.

\subsection{Magnetism in zero applied magnetic field}

High-$T_c$ superconductivity emerges from the gradual
destruction of the insulating AF parent compound by the
doping of charge carriers into the CuO$_2$ layers.
Shortly after their discovery, zero-field (ZF) $\mu$SR experiments 
on high-$T_c$ superconductors showed that static 
electronic moments are still present in lightly-doped
superconducting samples at low temperatures. In particular, 
an early ZF-$\mu$SR study of YBa$_2$Cu$_3$O$_y$ at
$T \! = \! 90$~mK showed that static magnetism was still 
present in samples with oxgen content $y \! \leq \! 6.51$, 
but not in a $y \! = \! 6.54$ sample \cite{Kiefl:89}.
Similarly, an early ZF-$\mu$SR study of 
La$_{2-x}$Sr$_x$CuO$_4$
\cite{Weidinger:89} reported static internal fields 
at low temperature for
$x \! \leq \! 0.15$. While these experiments suggested
that static magnetism and superconductivity coexist, there
was considerable worry that the samples studied contained
an inhomogeneous concentration of copper magnetic moments 
due to nonuniform doping of holes \cite{Heffner:89}.
Subsequent ZF-$\mu$SR experiments on pure La$_{2-x}$Sr$_x$CuO$_4$ 
indicated that static electronic moments are not present in samples 
with strontium content as high as $x \! = \! 0.15$ 
\cite{Kiefl:89,Niedermayer:98,Panagopoulos:02}, although the 
extrapolated $T \! \rightarrow \! 0$~K critical value of $x$ for the 
onset of static magnetism has never been accurately determined.
At dopings above where static magnetism is not observed in
ZF-$\mu$SR experiments, magnetic fluctuations persist that 
are visible by NMR and inelastic neutron scattering.

\begin{figure}
\includegraphics[width=10.0cm]{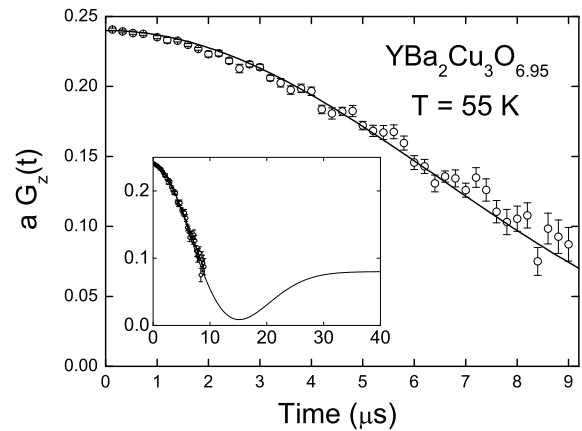}
\caption{Time evolution of the muon spin polarization in
YBa$_2$Cu$_3$O$_{6.95}$ at $T \! = \! 55$~K and $H \! = \! 0$,
measured with the initial polarization $P(0)$ perpendicular to
the $\hat{c}$-axis \cite{Sonier:01}. 
The solid curve is the static Gaussian
Kubo-Toyabe function with $\Delta \! = \! 0.1144$~$\mu$s$^{-1}$.
The inset shows what this function looks like beyond the time
range of the $\mu$SR signal. Note, the data shown here was 
obtained using a continuous-wave muon beam at TRIUMF (Vancouver,
Canada). With a pulsed muon source, such as that at ISIS (Oxford,
United Kingdom), it is very feasible to extend this $\mu$SR spectrum 
up to 20~$\mu$s.}   
\label{fig14}
\end{figure}

Whether the frozen spins observed by ZF-$\mu$SR in underdoped 
superconducting samples coexist with superconductivity or are 
simply due to phase segregated hole-poor regions, has been a 
central issue of controversy in the field. Establishing 
{\it coexistence} experimentally requires showing that a given 
sample exhibits both magnetism and superconductivity 
throughout 100~\% of its volume.
Since magnetic and non-magnetic regions in the same sample
give distinct $\mu$SR signals with amplitudes proportional to
the volume of the sample occupied by each phase,
in principle $\mu$SR can determine whether magnetism is present
in the entire sample. One constraint, however, is imposed by 
the sensitivity of $\mu$SR to nuclear dipole moments. Randomly orientated
nuclear dipoles cause a Gaussian relaxation of the ZF$\mu$SR time 
signal, as shown in figure \ref{fig14}. The solid curve
is a fit to a static Gaussian Kubo-Toyabe function
\begin{equation}
G_z^{\rm KT}(t) = \frac{1}{3} + \frac{2}{3}(1-\Delta^2 t^2) 
\exp ( -\frac{1}{2} \Delta^2 t^2) \, ,
\label{eq:KT}
\end{equation}
where $\Delta/\gamma_\mu$ is the width of the Gaussian field
distribution at the muon site due to the dense nuclear dipoles
(recall $\gamma_\mu \! = \! 0.0852$~$\mu$s$^{-1}$G$^{-1}$).
Typically, $\Delta \! \sim \! 0.1$~$\mu$s$^{-1}$, which corresponds
to a field of $\sim \! 1$~G at the muon site. A muon sitting
20~\AA~ away from a copper atom with a magnetic moment of $\sim \! 1$~$\mu_B$
will sense a $\sim \! 1$~G field, whereas a muon sitting further away will
not detect a field larger than that from the nuclear dipoles. Thus 
if all muons implanted in the sample experience a local magnetic field 
$> \! 1$~G, the nonmagnetic hole-rich regions cannot be larger
than $\sim \! 20$~\AA.
This conclusion was reached by Niedermayer {\it et al.} \cite{Niedermayer:98}
in ZF-$\mu$SR studies of the La$_{2-x}$Sr$_x$CuO$_4$ 
and Y$_{1-x}$Ca$_x$Ba$_2$Cu$_3$O$_6$ systems, 
where the hole doping in the CuO$_2$ layers is controlled by cation substitution.
Kanigel {\it et al.} \cite{Kanigel:02}
also concluded from ZF-$\mu$SR measurements that phase separation in
(Ca$_x$La$_{1-x}$)(Ba$_{1.75-x}$La$_{0.25+x}$)Cu$_3$O$_y$ is on
a microscopic scale. The maximum size of the nonmagnetic hole-rich 
regions inferred from these experiments is nearly equivalent to the
minimum possible spatial extent of the superconducting regions. The
latter is set by the superconducting coherence length $\xi_0$, which
is typically 15-20~\AA~ in high-$T_c$ superconductors.
Thus, the length scale of these ZF-$\mu$SR experiments is almost, but
not quite small enough to determine whether magnetism and superconductivity
coexist in the same spatial region.

\begin{figure*}
\includegraphics[width=18.0cm]{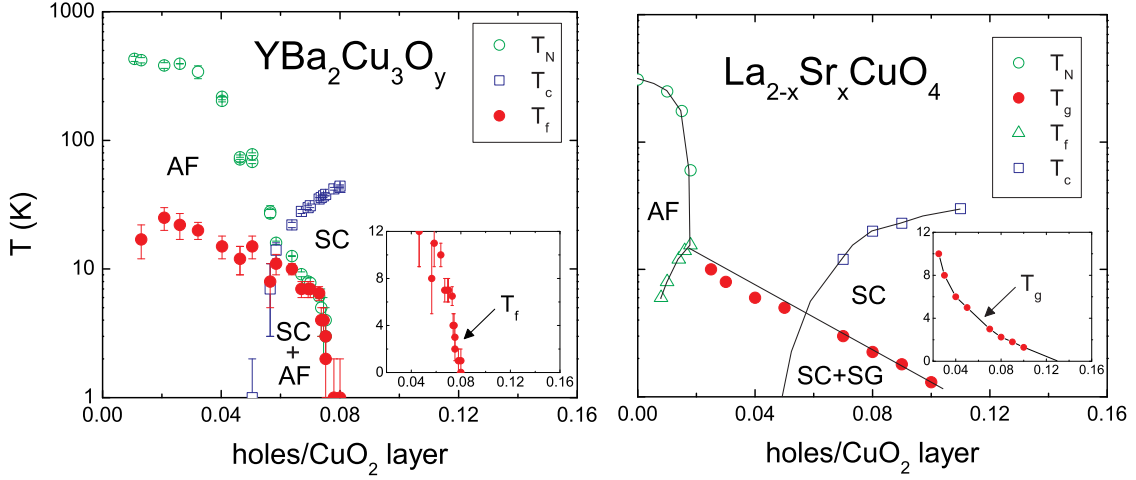}
\caption{Magnetic phase diagrams deduced from ZF-$\mu$SR studies of 
YBa$_2$Cu$_3$O$_y$ \cite{Sanna:04} and La$_{2-x}$Sr$_x$CuO$_4$ \cite{Niedermayer:98}.
In YBa$_2$Cu$_3$O$_y$, static short-range AF correlations are detected
below the spin-freezing temperature $T_f$. In La$_{2-x}$Sr$_x$CuO$_4$, 
a `cluster' spin-glass (SG) state appears below $T_g$. In both materials there
is a microscopic coexistence of superconductivity (SC) and frozen 
AF correlations in underdoped samples. The insets in each panel show the
$T \! \rightarrow \! 0$~K extrapolation of the transition temperatures
for static magnetism plotted on a linear scale.}
\label{fig15}
\end{figure*} 

While the ZF-$\mu$SR studies of Refs.~\cite{Niedermayer:98,Kanigel:02}
provided strong evidence for the coexistence of magnetism and superconductivity
on a nano-scale level, a couple of questions about these works have lingered.
The first is whether superconductivity occurs in the entire
bulk of the sample. The second question is whether the inherent cation disorder 
that is present in the systems studied in Refs.~\cite{Niedermayer:98,Kanigel:02}
is responsible for the occurrence of electronic moments in 
superconducting samples.
To address these questions, more recently, ZF-$\mu$SR, TF-$\mu$SR and bulk 
magnetization measurements were 
performed on both high-quality polycrystalline samples \cite{Sanna:04}
and single crystals \cite{Miller:06} of YBa$_2$Cu$_3$O$_y$.
In contrast to the systems studied in Refs.~\cite{Niedermayer:98,Kanigel:02},
hole doping in YBa$_2$Cu$_3$O$_y$ is done by varying the oxygen concentration
in the CuO chain layers, rather than by cation substitution.
While superconductivity is widely believed to take place within the CuO$_2$ 
planes, the chains themselves are conductive and superconducting 
\cite{Basov:95,Lee:05}, due to strong coupling with the nearby CuO$_2$ planes 
\cite{Atkinson:95,Xiang:96}. Furthermore, charge ordering in the CuO chain 
layers induces charge-density modulations in the CuO$_2$ layers
\cite{Grevin:00}. Precision ZF-$\mu$SR measurements on YBa$_2$Cu$_3$O$_y$
show that the charge ordering affects the relaxation of the $\mu$SR signal
\cite{Sonier:02}. However, this could be due to a change in the nuclear
dipole field sensed by the muon, rather than the formation of electronic
moments. More importantly, the effects of charge ordering on the ZF-$\mu$SR
spectrum weakens with decreasing oxygen content, making it easy to identify
the onset of static electronic moments in lightly doped samples.

The more recent ZF-$\mu$SR of studies of high-quality 
YBa$_2$Cu$_3$O$_y$ samples \cite{Sanna:04,Miller:06} are consistent with the
earlier experiments, in that static electronic moments occurring
in superconducting samples with a hole-doping concentration less than
$p \! \approx \! 0.08$ are sensed by all implanted muons. 
In samples where $T_c$ is greater than
the onset temperature for static electronic moments, 
TF-$\mu$SR measurements above the spin freezing temperature $T_f$ 
(or $T_M$ in Ref.~\cite{Miller:06}) show that a vortex lattice is formed 
throughout the bulk of the sample. In other words, the entire sample
exhibits superconductivity. Below $T_f$, the same conclusion cannot be
reached, because the magnetism rapidly depolarizes the TF-$\mu$SR signal.    
However, bulk magnetization measurements show no reduction in the
superconducting response below $T_f$. Thus, the experiments of
Refs.~\cite{Sanna:04,Miller:06} appear to establish that static 
magnetism and superconductivity coexist on a microscopic scale. Saying
any more than this goes beyond the spatial resolution of the ZF-$\mu$SR
experiments.

The arrangement of the frozen Cu spins in the coexistence phase
inferred from the ZF-$\mu$SR experiments may be compared to that
indicated by neutron scattering experiments. The ZF-$\mu$SR experiments 
on La$_{2-x}$Sr$_x$CuO$_4$ 
\cite{Niedermayer:98} show that a cluster spin-glass (CSG) phase
persists in the superconducting state and is characterized by an onset 
temperature that decreases continuously across the insulator-superconducting 
boundary. The CSG phase, which has also been detected by $^{139}$La 
nuclear quadrupole resonance (NQR) \cite{Cho:92,Chou:93,Julien:99}, 
consists of AF spin 
correlations within domains that are separated by walls of hole-rich
material, with the easy axis of the staggered magnetization uncorrelated 
between clusters. Neutron scattering experiments on La$_{2-x}$Sr$_x$CuO$_4$
show that both
static and dynamic incommensurate spin correlations exist across the       
insulator-superconducting boundary \cite{Birgeneau:06}. The source 
of the low-temperature {\it static} incommensurate spin correlations that 
give rise to an elastic neutron scattering peak shifted with respect to the 
AF position for $0.2 \! \leq \! x \! \leq \! 0.12$ samples
\cite{Wakimoto:99,Matsuda:00,Wakimoto:01}, is likely the same static
(or quasistatic) magnetism detected by the low-frequency $\mu$SR and 
NQR measurements. In fact, Wakimoto {\it et al.} 
\cite{Wakimoto:01} showed that the ordered magnetic moment decreases 
continuously across the insulator-superconducting boundary.
In recent years, spin-spiral models \cite{Shraiman:89} have emerged
to explain the magnetic incommensurability of 
La$_{2-x}$Sr$_x$CuO$_4$. 
L\"{u}scher {\it et al.} \cite{Luscher:07} have proposed that the CSG 
phase observed at low $T$ by ZF-$\mu$SR results from the freezing of 
incompatible spin-spiral configurations in neighbouring CuO$_2$ layers.       

\begin{figure*}
\includegraphics[width=18.0cm]{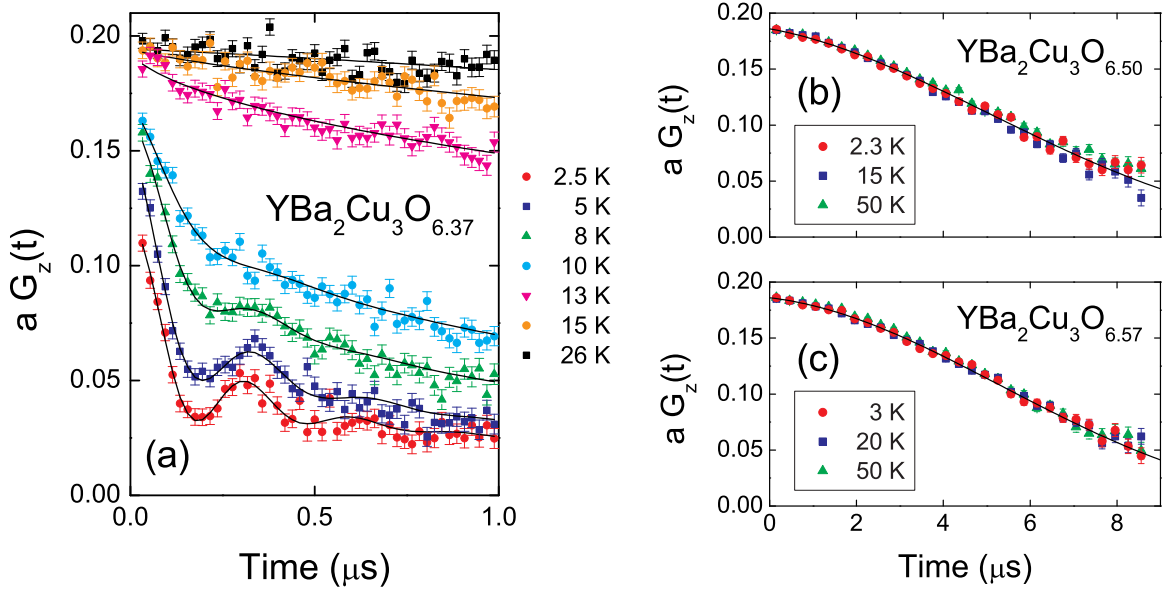}
\caption{Temperature dependence of the ZF-$\mu$SR time signal for
YBa$_2$Cu$_3$O$_y$
singles crystals recorded with the initial muon-spin polarization 
$P(0)$ parallel to the $\hat{c}$-axis. The spectra for 
$y \! = \! 6.37$ come from the measurements of 
Ref.~\cite{Miller:06}, whereas the spectra for $y \! = \! 6.50$
and $y \! = \! 6.57$ come from the work of Ref.~\cite{Sonier:06b}.
The oscillations observed at low $T$ for $y \! = \! 6.37$ indicate
short-range magnetic order. The ZF-$\mu$SR signals of 
$y \! = \! 6.50$ and $y \! = \! 6.57$ are identical,
temperature-independent, and described by relaxation due to 
nuclear dipole moments and an additional weak exponential relaxation
of unknown origin \cite{Sonier:01}.}   
\label{fig16}
\end{figure*}

\begin{figure*}
\includegraphics[width=16.0cm]{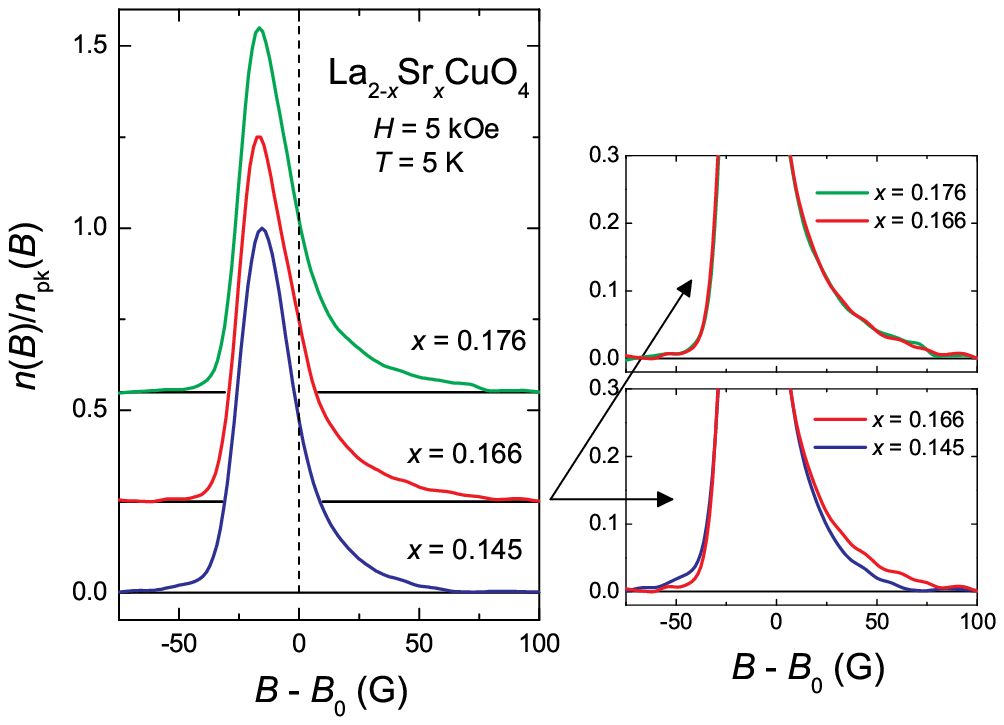}
\caption{$\mu$SR line shapes for La$_{2-x}$Sr$_x$CuO$_4$ single crystals
at $H \! = \! 5$~kOe and $T \! = \! 5$~K. 
For comparison, all line shapes have been 
normalized to their respective peak amplitude $n_{\rm pk}(B)$.
The dashed vertical line indicates the applied field $B_0$. The
background signal from muons missing the sample and stopping
elsewhere is very small, and hence there is no visible peak
at $B \! = \! B_0$. The two panels on the right
are blowups of the lower portion of the $\mu$SR line shapes,
with the widths of the line shapes made equivalent to account
for changes in $\lambda_{ab}$. The shape of $n(B)$ clearly 
changes between $0.145 \! < \! x \! < \! 0.166$.}
\label{fig17}
\end{figure*}

The ZF-$\mu$SR signals for $p \! < \! 0.08$ superconducting samples of 
YBa$_2$Cu$_3$O$_y$ at low temperature are characterized by a rapidly damped 
oscillation at early times \cite{Kiefl:89,Sanna:04,Miller:06}. An example
of this is shown in figure~\ref{fig16} for YBa$_2$Cu$_3$O$_{6.37}$.
The rapidly damped oscillation is indicative of very short-range magnetic 
order. Sanna {\it et al.} \cite{Sanna:04} have argued that this is consistent
with nanoscale coexistence of stripe-like AF domains and superconducting
material. The situation is in contrast to the coexistence of CSG magnetism 
and superconductivity observed in cation-substituted cuprates, which
could be argued to be an extrinsic property arising from inherent disorder
\cite{Sanna:05}. Recent neutron scattering measurements by
Stock {\it et al.} \cite{Stock:06} on YBa$_2$Cu$_3$O$_{6.353}$ single crystals 
support the occurrence of a single phase of coexisting superconductivity 
and short-range spin correlations. Fluctuating short-range commensurate
spin correlations that develop at temperatures well above $T_c$ gradually 
slow down to form a spin glass-like state at low temperature. 
However, in contrast to the ZF-$\mu$SR measurements, no transition to 
magnetic order of any kind is seen in the neutron measurements.
Stock {\it et al.} have suggested that the magnetic freezing transition 
temperature observed in the ZF-$\mu$SR experiments of 
Refs.~\cite{Sanna:04,Miller:06} is simply the temperature at which the 
slow spin dynamics observed by neutron scattering comes into the 
$\mu$SR time window. While this explanation is reasonable, it does not 
account for the oscillation observed in the ZF-$\mu$SR signal at low
$T$. A combined $\mu$SR and neutron scattering study on the same 
severely underdoped sample may reconcile this discrepancy.

\subsection{Field-induced/enhanced magnetism}
 
NMR experiments on near-optimally doped 
YBa$_2$Cu$_3$O$_{y}$ \cite{Curro:00,Mitrovic:01,Mitrovic:03}, 
YBa$_2$Cu$_4$O$_8$ \cite{Kakuyanagi:02} and
Tl$_2$Ba$_2$CuO$_{6 + \delta}$ \cite{Kakuyanagi:03} have detected AF
spin fluctuations in the vicinity of the vortex cores. Likewise,  
low-energy field-induced spin fluctuations have been detected by inelastic 
neutron scattering experiments on La$_{1.837}$Sr$_{0.163}$CuO$_4$
\cite{Lake:01}. As the fast spin fluctuation rates detected in these 
experiments fall outside the $\mu$SR time window, the field-induced
spin dynamics in the high-doping regime cannot be studied by $\mu$SR.    
On the other hand, spin fluctuations slow down considerably in the
low-doped regime and it is here where further information has recently
been obtained by $\mu$SR.

\subsubsection{La$_{2-x}$Sr$_x$CuO$_4$}

\begin{figure}
\includegraphics[width=10.0cm]{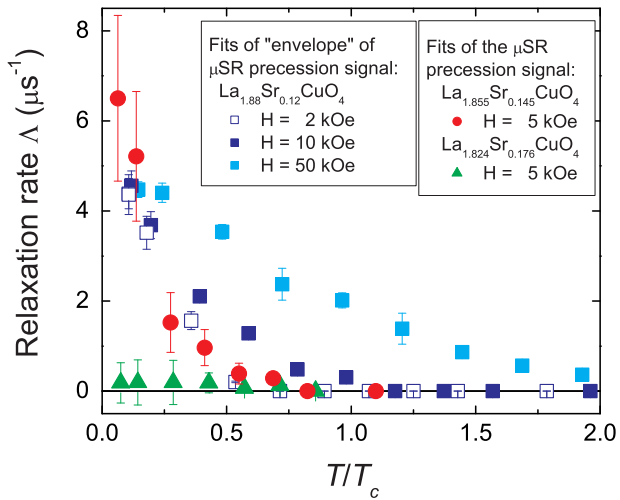}
\caption{Temperature dependence of the exponential 
relaxation rate $\Lambda$ from
fits of the TF-$\mu$SR signal for La$_{1.855}$Sr$_{0.145}$CuO$_4$ 
at $H \! = \! 5$~kOe (circles) and La$_{1.824}$Sr$_{0.176}$CuO$_4$ 
at $H \! = \! 5$~kOe (triangles) to equation~(\ref{eq:polSG})
\cite{Sonier:06b}. Temperature dependence of the exponential relaxation 
rate $\Lambda_1$ (squares) from fits of the `envelope' of the
TF-$\mu$SR signal for La$_{1.88}$Sr$_{0.12}$CuO$_4$ at $H \! = \! 2$~kOe,
10~kOe and 50~kOe \cite{Savici:05}.
Note, $\Lambda$ from Ref.~\cite{Sonier:06b} is associated with
magnetism at the centre of the vortex core ($r \! = \!0$),
where superconductivity is fully suppressed.}    
\label{fig18}
\end{figure}

There have been a couple of TF-$\mu$SR studies providing evidence for frozen spins in 
the vortex cores of La$_{2-x}$Sr$_x$CuO$_4$. Kadono {\it et al.} \cite{Kadono:04a} 
investigated the temperature dependence of the vortex core size at $H \! = \! 2$~kOe
in both underdoped and overdoped La$_{2-x}$Sr$_x$CuO$_4$ single crystals. 
While the temperature dependence
of the vortex core size at $x \! = \! 0.19$ was found to be in qualitative 
agreement with the Kramer-Pesch effect, in the $x \! = \! 0.13$ and $x \! = \! 0.15$ 
samples the vortex core size increased with decreasing temperature and to 
a value significantly larger than expected from measurements of $H_{c2}$. 
The large size of the vortex cores in the lower doped samples was 
argued to result from a suppression of superconductivity around the vortex
due to the formation of AF spin correlations. At the time, this seemed compatible
with neutron experiments on La$_{2-x}$Sr$_x$CuO$_4$ \cite{Lake:02,Katano:00} showing 
enhanced static AF spin correlations in underdoped samples. However, 
Kadono {\it et al.} speculated that the spin correlations responsible for 
the large vortex core size at low $T$ and
low $H$ in La$_{2-x}$Sr$_x$CuO$_4$ with $x \! = \! 0.13$ and $x \! = \! 0.15$ 
are not static, but dynamic. 
This interpretation was supported by a subsequent neutron
study of an $x \! = \! 0.144$ sample 
by Khaykovich {\it et al.} \cite{Khaykovich:05}
that revealed static magnetic order only above $H \! \approx \! 30$~kOe.
However, not ruled out by these studies is the possibility that there is
disordered static magnetism at low fields.   
  
The latter possibility was considered in a more recent $\mu$SR study
of La$_{2-x}$Sr$_x$CuO$_4$ single crystals that are 
free of static magnetism in zero field 
\cite{Sonier:06b}. Visually, evidence for static magnetism 
confined to the vortex cores can be seen in the TF-$\mu$SR line shape as an 
unusual high-field `tail' and the possible appearance of a low-field tail 
(depending on the magnitude and orientation of the fields in the vortex core), 
but with no other change in the shape of $n(B)$.
As shown in figure \ref{fig17}, 
the high-field tail of the $\mu$SR line shape for La$_{2-x}$Sr$_x$CuO$_4$ changes
at a Sr doping $0.145 \! \leq \! x \! \leq 0.166$. However, similar changes 
of the $\mu$SR line shape could also arise from a 3D-to-2D crossover of the
FLL. For example, in highly anisotropic
Bi$_{2+x}$Sr$_{2-x}$CaCu$_2$O$_{8+\delta}$ Lee {\it et al.} \cite{Lee:93}
showed that above a crossover field, random pinning-induced misalignment of 
2D `pancake' vortices in adjacent layers narrows and 
reduces the asymmetry of the $\mu$SR
line shape. For La$_{2-x}$Sr$_x$CuO$_4$, a similar scenario 
would imply that a reduction
in Sr substitution significantly softens the flux lines and induces
a degree of disorder sufficient to misalign the 2D vortices in adjacent
layers. However, a combined $\mu$SR and neutron study of
La$_{1.90}$Sr$_{0.10}$CuO$_4$ by Divakar {\it et al.} \cite{Divakar:04} 
demonstrated that the vortices are 3D-like even at this 
lower Sr concentration. This conclusion is supported by neutron 
scattering experiments on La$_{1.90}$Sr$_{0.10}$CuO$_4$ by
Lake {\it et al.} \cite{Lake:05}, which show that the field-induced 
static magnetic order is 3D.

While the onset of magnetism is a more likely cause of the change in the 
$\mu$SR line shape of La$_{2-x}$Sr$_x$CuO$_4$ with doping, 
the neutron study of Khaykovich {\it et al.} \cite{Khaykovich:05} 
implies that the line shape for $x \! = \! 0.145$ measured at 
low $T$ and low $H$ in Ref.~\cite{Sonier:06b} does not result from static 
magnetic order. However, the TF-$\mu$SR precession signals were found to
be well described by the following polarization function, which allows for 
spin-glass like magnetism in and around the vortex cores
\begin{equation}
P(t) = \sum_i \exp(-\Lambda e^{-(r_i/\xi_{ab})^2} t) \cos[\gamma_\mu B(r_i) t].\
\label{eq:polSG}  
\end{equation} 
Here the sum is over all sites in the real-space unit cell of the FLL and $B(r_i)$ 
is the local field at position $r_i \! = \! (x_i, y _i)$ with respect to the vortex 
center. The distribution of fields at each site is assumed to be 
Lorentzian, and the relaxation rate $\Lambda$ due to the magnetism is assumed to 
fall off as a function of radial distance from the vortex core centre on the scale 
of $\xi_{ab}$.
The temperature dependences of $\Lambda$ for La$_{2-x}$Sr$_x$CuO$_4$ 
with $x \! = \! 0.145$ and $x \! = \! 0.16$ at 
$H \! = \! 5$~kOe are shown in figure \ref{fig18}.
The increase of $\Lambda$ with decreasing temperature observed for   
the $x \! = \! 0.145$ sample indicates a slowing down of spin fluctuations
in the vortex-core region.
 
\begin{figure*}
\includegraphics[width=16.0cm]{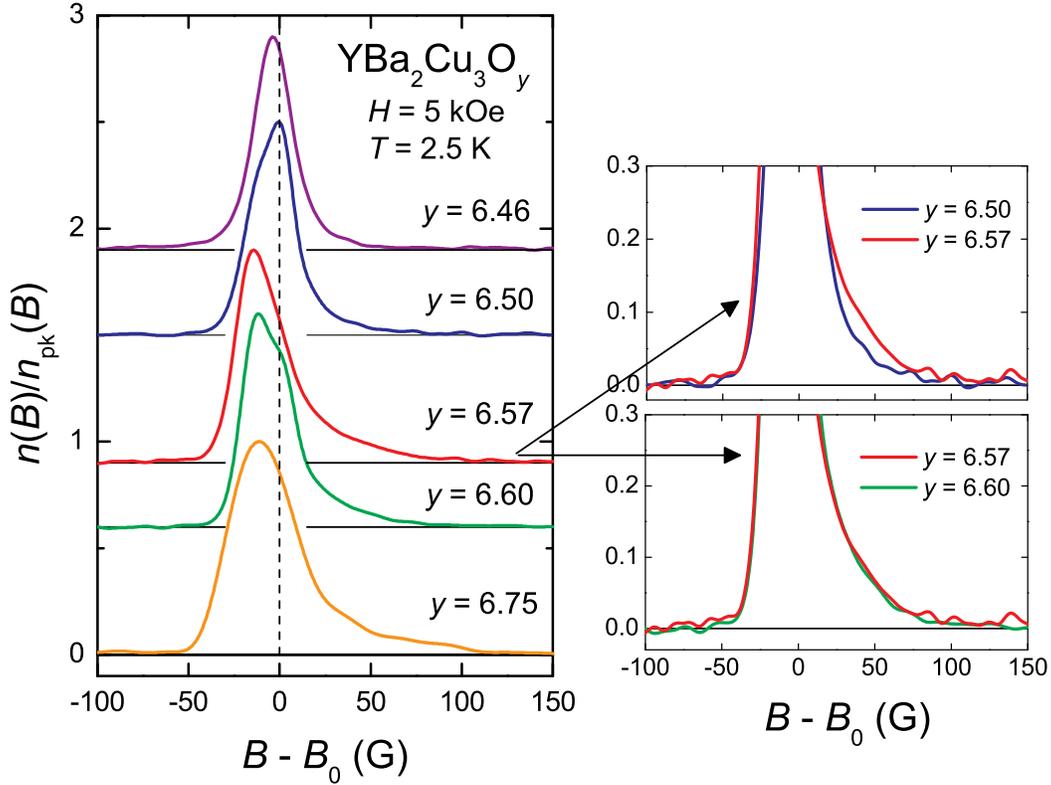}
\caption{$\mu$SR line shapes for YBa$_2$Cu$_3$O$_y$ single crystals
at $H \! = \! 5$~kOe and $T \! = \! 2.5$~K. 
For comparison, all line shapes have been 
normalized to their respective peak amplitude $n_{\rm pk}(B)$.
The dashed vertical line indicates the applied field $B_0$, and the
bump or peak near this line (most visible for $y \! = \! 6.50$)
is the background signal. The two panels on the right
show blowups of the lower portion of the $\mu$SR line shapes,
with the widths of the line shapes made equivalent to account
for changes in $\lambda_{ab}$. The shape of $n(B)$ clearly 
changes between $6.50 \! < \! y \! < \! 6.57$.}
\label{fig19}
\end{figure*}

It is important to emphasize that these fits to the TF-$\mu$SR time
spectra don't in themselves prove the existence of spin-glass order
in the vortex cores. Other models based on different microscopic physics,
including different forms of magnetism,
may give equally good or better fits to the time-domain signals.
The conclusion that magnetism occurs in the vortex cores really comes
from the changes in the TF-$\mu$SR line shape that are visually
apparent as a function of temperature, magnetic field and hole doping. 
When the changes as a function of these three independent variables 
are considered together, there are really only two logical interpretations 
of the data. As already mentioned, experiments that
have probed the dimensionality of the vortices in La$_{2-x}$Sr$_x$CuO$_4$
more or less rule out the cause being a vortex transition. 
In the next section, experimental evidence that also 
argues against this being the interpretation for similar changes of the
TF-$\mu$SR line shape of YBa$_2$Cu$_3$O$_y$ is discussed. 

With this in mind, the results of the fits for La$_{1.855}$Sr$_{0.145}$CuO$_4$  
may be compared to recent TF-$\mu$SR measurements of
La$_{1.88}$Sr$_{0.12}$CuO$_4$ by Savici {\it et al.} \cite{Savici:05}, showing a 
field-induced enhancement of static or quasistatic magnetism. 
In contrast to the TF-$\mu$SR studies discussed so far,
information on field-induced magnetism was obtained in this work by
fitting the `envelope' of the muon spin precession signal at early
times to a two-component exponential function 
$E_1 \exp(-\Lambda_1 t) + E_2 \exp(-\Lambda_2 t)$, where the first and
second terms describe contributions to the $\mu$SR signal from
volume fractions of the sample with and without quasistatic magnetism, 
respectively. This was possible for La$_{1.88}$Sr$_{0.12}$CuO$_4$, 
because the field-enhanced magnetism occupies a volume of the sample much 
larger than the volume of the vortex cores. In particular, 
$E_1/E_2 \! > \! 60$~\% compared to $E_1/E_2 \! < \! 1$~\%
La$_{1.855}$Sr$_{0.145}$CuO$_4$ in Ref.~\cite{Sonier:06b}.  
As shown in figure \ref{fig18}, the temperature dependence of 
the relaxation rate $\Lambda$ of La$_{1.855}$Sr$_{0.145}$CuO$_4$ determined
from fits of the full TF-$\mu$SR precession signal with 
equation~(\ref{eq:polSG}) is similar to the
temperature dependence of the relaxation rate
$\Lambda_1$ determined from fits to the envelope function of 
La$_{1.88}$Sr$_{0.12}$CuO$_4$ at low $H$.
This seems to suggest that the static (or quasistatic) magnetism that 
occurs in a large volume of La$_{1.88}$Sr$_{0.12}$CuO$_4$ in a magnetic
field is similar to the magnetism found near the vortex cores in 
La$_{1.855}$Sr$_{0.145}$CuO$_4$. However, this is unlikely the case 
at low temperatures. In contrast to La$_{1.855}$Sr$_{0.145}$CuO$_4$,
static magnetic correlations occur in La$_{1.88}$Sr$_{0.12}$CuO$_4$ even 
in zero field \cite{Katano:00}. Thus the fast relaxation rate observed in 
La$_{1.88}$Sr$_{0.12}$CuO$_4$ in the vortex state at low $T$ can be
attributed to static magnetic order.  

The study by Savici {\it et al.} also  
showed that quasistatic magnetism persists to temperatures well above 
$T_c$ at high $H$. This implies that vortices are not required to nucleate 
static magnetism in superconducting samples. Instead the vortices
act to suppress spin fluctuations and enhance the magnetic volume 
fraction. The TF-$\mu$SR experiments by Savici {\it et al.} \cite{Savici:05}
provide strong evidence for these effects, as well as neutron scattering 
experiments on La$_{1.88}$Sr$_{0.12}$CuO$_4$ by Katano {\it et al.} \cite{Katano:00} 
showing field-enhanced AF spin correlations at $H \! = \! 100$~kOe
and low $T$, and Raman scattering experiments on La$_{1.88}$Sr$_{0.12}$CuO$_4$
by Machtoub {\it et al.} \cite{Machtoub:05} that can be explained in terms of
a field-enhanced AF volume fraction.   
     
\subsubsection{YBa$_2$Cu$_3$O$_y$.}

Miller {\it et al.} \cite{Miller:02} reported evidence from TF-$\mu$SR 
measurements on YBa$_2$Cu$_3$O$_{6.50}$ at $H \! \geq \! 10$~kOe for 
the occurrence of static AF order in the vortex cores. 
While Kadono 
{\it et al.} \cite{Kadono:05} were able to fit low field TF-$\mu$SR spectra 
for underdoped La$_{2-x}$Sr$_x$CuO$_4$ 
assuming the spatial field profile $B({\bf r})$
of the FLL is adequately described by the London model with a 
Lorentzian cutoff factor $F(G) \! = \! \exp(- \sqrt{2} G \xi)$, 
fits of the $\mu$SR time spectra for 
YBa$_2$Cu$_3$O$_{6.50}$ assuming a conventional London or 
GL model were found not to converge.
In Ref.~\cite{Miller:02} the TF-$\mu$SR signal from YBa$_2$Cu$_3$O$_{6.50}$ 
could be fit assuming static AF order in the vortex cores commensurate with 
the underlying crystal lattice, even though
to date field-induced static magnetic order has not been detected 
in YBa$_2$Cu$_3$O$_y$ by neutron scattering. 
The magnitude of the 
staggered field sensed at the 
{\it muon stopping site(s)} was determined to be $\sim \! 18$~G at low $T$. 
However, this analysis still required the value of $\xi_{ab}$ to 
be constrained, which suggests an insensitivity to the high-field tail of
the $\mu$SR line shape or that the assumed theoretical model is
inappropriate. 

Recently, the doping, temperature and magnetic field dependences of the $\mu$SR 
line shape for YBa$_2$Cu$_3$O$_y$ single crystals 
were studied in some detail \cite{Sonier:06b}. 
$\mu$SR line shapes for YBa$_2$Cu$_3$O$_{6.46}$ and YBa$_2$Cu$_3$O$_{6.50}$ were 
found to be compatible with spin-glass magnetism appearing in and around 
the vortex cores (some of these measurements are shown in figure~\ref{fig19}).
This interpretation of the $\mu$SR signal explains
why frozen spins have not been detected in the vortex cores of YBa$_2$Cu$_3$O$_y$
by neutron scattering, which is insensitive to static magnetism with random spatial
correlations.
As in the case of La$_{2-x}$Sr$_x$CuO$_4$, an explanation of the observed 
changes in the TF-$\mu$SR line shape in terms of a crossover in the 
dimensionality of the vortices is unlikely. 
Josephson plasma resonance measurements by Dulic {\it et al.} \cite{Dulic:01}
indicate that the vortices in YBa$_2$Cu$_3$O$_{6.50}$ are 3D-like at 
low temperatures. Consistent with this finding, mutual inductance measurements 
show that even severely underdoped YBa$_2$Cu$_3$O$_y$ becomes quasi-2D
only near $T_c$ \cite{Zuev:05}. 
     
\subsubsection{Pr$_{2-x}$Ce$_x$CuO$_4$ and Pr$_{1-x}$LaCe$_x$CuO$_4$.}

\begin{figure}
\includegraphics[width=12.0cm]{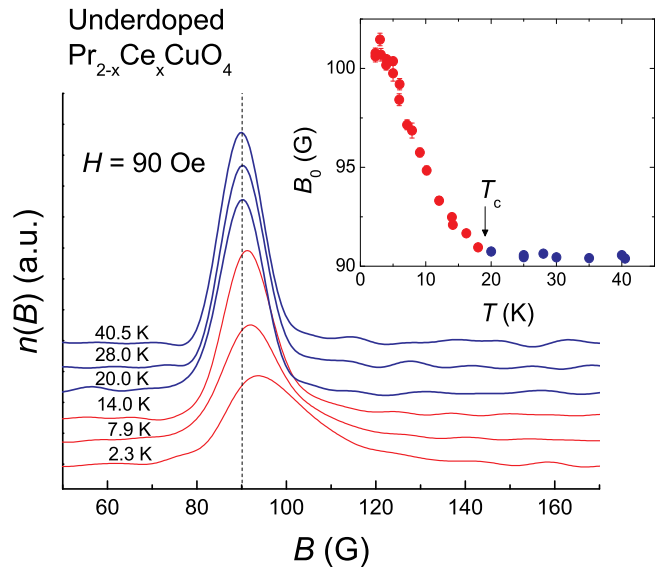}
\caption{Temperature dependence of the $\mu$SR line shape in
an underdoped Pr$_{2-x}$Ce$_x$CuO$_4$ single crystal at
$H \! = \! 90$~Oe \cite{Sonier:03}. Inset: Temperature
dependence of the average internal magnetic field $B_0$.
The blue and red colour scheme denotes measurements taken above and
below $T_c$, respectively.}
\label{fig20}
\end{figure}

\begin{figure*}
\includegraphics[width=19.0cm]{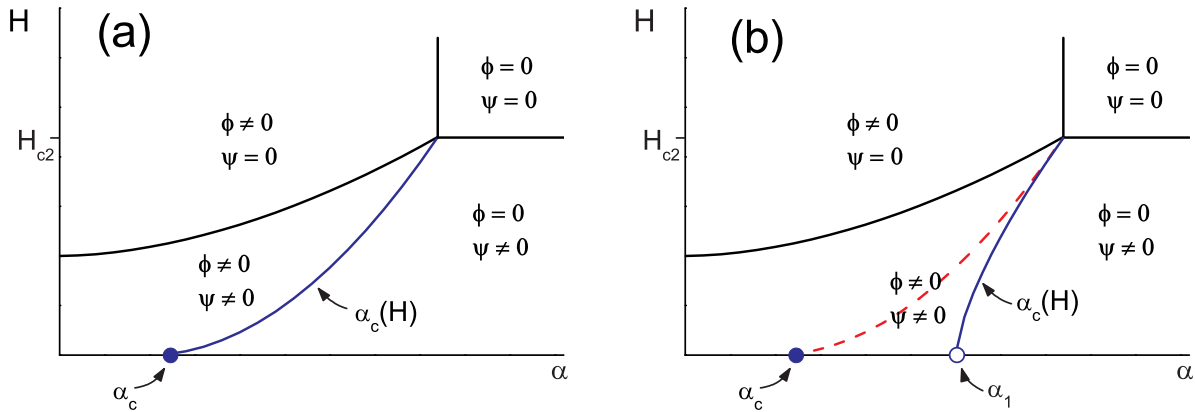}
\caption{Schematic of theoretically predicted zero-temperature 
phase diagrams for (a) 2D vortices 
(adapted from figure~1 of Ref.~\cite{Demler:01}), 
and (b) 3D vortices (adapted from figure~1 of Ref.~\cite{Kivelson:02}).
Here $\alpha$ is a control parameter ({\it e.g.} carrier 
concentration). $\Psi$ and $\phi$ denote the expectation values of 
the superconducting and competing order parameters, respectively. 
The solid blue curve represents a quantum phase transition (QPT) and
the solid blue dot denoted $\alpha_c$ is a quantum critical point 
(QCP). Note that the QPT for the 2D vortex case is a crossover within the
coexistence phase of the 3D vortex phase diagram (red dashed curve).
Below the crossover curve, the competing order is spatially
inhomogeneous. In addition, the zero field extrapolation of the 
QPT in (b) is an `avoided' QCP (open blue circle).}
\label{fig21}
\end{figure*}

The first reported experimental evidence for field-induced magnetic order
in an electron-doped cuprate superconductor came from a TF-$\mu$SR study
of underdoped Pr$_{2-x}$Ce$_x$CuO$_4$ single crystals \cite{Sonier:03}.
In zero field, random magnetism was detected, but 
a remarkably weak applied field of only 90~Oe resulted in the occurrence
of static AF order thoughout the volume of the sample. The onset of bulk AF
order below $T_c$ is visually apparent from the observed positive
field-shift of the {\it entire} $\mu$SR line shape (see figure \ref{fig20}).
In this case, the entire line shape shifts to higher $B$ because
the AF order enhances the local magnetic field at all of the O(3)
muon stopping sites. 

The occurrence of weak field-induced AF order was
subsequently identified in superconducting Pr$_{1-x}$LaCe$_x$CuO$_4$ 
with $x \! = \! 0.11$ and $x \! = \! 0.15$,  
in TF-$\mu$SR studies by Kadono {\it et al.} \cite{Kadono:04b,Kadono:05}.
In these experiments it was argued that the muon senses the 
antiferromagnetism indirectly through a transferred hyperfine coupling
between the $\mu^+$ spin and the Cu moments via the Pr ions, rather
than by direct detection of the dipolar fields of the Cu moments at
the muon site as suggested in Ref.~\cite{Sonier:03,Sonier:04b}. 

Single crystals of Pr$_{1-x}$LaCe$_x$CuO$_4$ 
can be made much larger than single crystals 
of Pr$_{2-x}$Ce$_x$CuO$_4$, and this has allowed neutron scattering 
studies of this compound. Fujita {\it et al.} used neutrons to study the same
Pr$_{1-x}$LaCe$_x$CuO$_4$ samples that 
Kadono {\it et al.} investigated with $\mu$SR \cite{Fujita:04}.
In zero field, weak AF order was detected in the $x \! = \! 0.11$ sample,
in contrast to ZF-$\mu$SR measurements that found only random
magnetism due to small Pr moments \cite{Kadono:03}.
This apparent discrepancy suggests that the AF correlations 
in zero field are fluctuating on the $\mu$SR time scale. The neutron experiments
by Fujita {\it et al.} also demonstrated a field-induced enhancement of 
AF order in the $x \! = \! 0.11$ sample, whereas static AF order
was not observed at all in the $x \! = \! 0.15$ sample. The latter
result contrasts the TF-$\mu$SR experiments of Kadono {\it et al.} who
reported field-induced AF order in the $x \! = \! 0.15$ sample, 
but with a reduced ordered local moment.
More recently, Kang {\it et al.} have detected static SDW order and 
residual AF order in underdoped Pr$_{1-x}$LaCe$_x$CuO$_4$ at zero field 
by neutron scattering, but not in optimally doped 
Pr$_{1-x}$LaCe$_x$CuO$_4$ \cite{Kang:05}. Furthermore, they
showed that a $\hat{c}$-axis aligned magnetic field enhances 
the SDW order, but not the AF order in underdoped 
Pr$_{1-x}$LaCe$_x$CuO$_4$, and does 
not induce static magnetic order of either kind in optimally doped 
Pr$_{1-x}$LaCe$_x$CuO$_4$.  

\subsection{Piecing together the information from $\mu$SR}

ZF-$\mu$SR and TF-$\mu$SR studies of cuprates have established that
superconductivity and static (quasistatic) magnetism coexist
on the low-doping side of the superconducting `dome'. 
In quality samples, this coexistence is on a 
nanometer scale at zero magnetic field. {\it Field-induced} static 
magnetism has been detected at low temperatures in samples where the 
spin fluctuation rate at $H \! = \! 0$ falls outside the $\mu$SR time 
window. In these samples, disordered static magnetism is induced 
by weak magnetic fields. 

As discussed in Ref.~\cite{Birgeneau:06}, neutron scattering studies 
of La$_{2-x}$Sr$_x$CuO$_4$ and La$_2$CuO$_{4+y}$ support a theoretical
model of competing superconducting and magnetic order parameters
proposed by Demler {\it et al.} \cite{Demler:01}.
In this model the pure superconductor undergoes a field-dependent 
quantum phase transition (QPT) to a state of coexisting superconductivity
and magnetic order (see figure~\ref{fig21}a). 
The competing magnetic order is stabilized at the 
QPT by the suppression of superconductivity associated with
the formation of vortices.        
In agreement with the theory, the neutron experiments
have detected {\it field-induced} magnetic order \cite{Khaykovich:05} and  
have shown that the ordered magnetic moment grows with increasing magnetic 
field exactly as predicted \cite{Lake:02,Khaykovich:02}.
The theoretical model of Demler {\it et al.} is not exclusive to the
cuprates, as it also appears to describe the occurrence of 
a field-induced coexistence phase of magnetic order and superconductivity
in CeRhIn$_5$ \cite{Park:06}. 

In actuality the theoretical model of Demler {\it et al.} is 
somewhat oversimplified. In particular, the model assumes that the 
vortices are 2D. As a finite size system, an isolated 2D vortex
cannot support static magnetic order. It is only when the 
2D vortices strongly couple to their neighbours is static magnetic order 
stabilized. Thus the coexistence phase in the theory of Demler {\it et al.}
is characterized by a competing order parameter that is nearly
spatially uniform throughout the sample. However, as mentioned earler,
neutron scattering studies of La$_{1.90}$Sr$_{0.10}$CuO$_4$ by 
Lake {\it et al.} \cite{Lake:05} show that the magnetic order in 
the vortex state is in fact 3D.
As explained in Ref.~\cite{Lake:05}, this implies that there is 
significant coupling between the CuO$_2$ planes, so that the vortices 
themselves are 3D. As discussed earlier, this same conclusion was
reached in a TF-$\mu$SR study of La$_{1.90}$Sr$_{0.10}$CuO$_4$
by Divakar {\it et al.} \cite{Divakar:04}.

\begin{figure*}
\includegraphics[width=16.0cm]{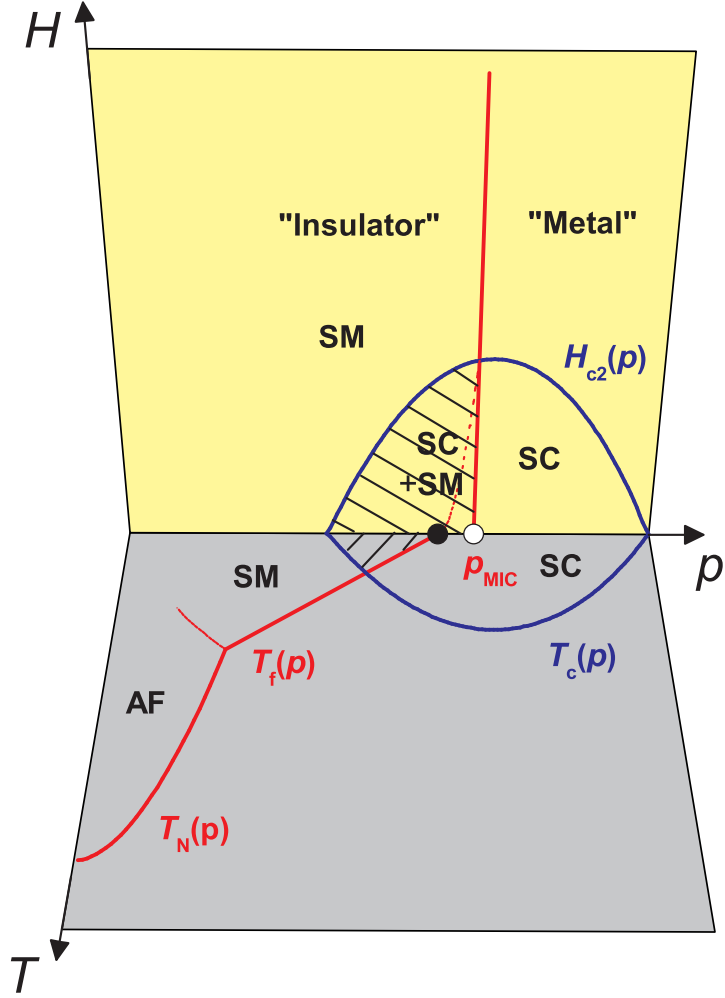}
\caption{Generic phase diagram inferred from $\mu$SR studies
on high-$T_c$ superconductors. 
The acronyms for the different phases are defined as
follows: {\bf AF} $\equiv$ antiferromagnetic phase,
{\bf SM} $\equiv$ static magnetic phase (where magnetism 
may be disordered), {\bf SC} $\equiv$ superconducting phase,
and {\bf SC + SM} $\equiv$ coexisting superconducting and static
magnetic phases. 
$T_c(p)$, $T_{\rm N}(p)$, $T_f(p)$, $H_{c2}(p)$ and $p_{\rm MIC}$ 
denote the superconducting critical temperature, the Neel temperature, 
the spin freezing temperature, the
upper critical magnetic field and the $T \! \rightarrow \! 0$ 
metal-to-insulator crossover (MIC), respectively. 
TF-$\mu$SR measurements indicate that the crossover at 
$p_{\rm MIC}$ extends below $H_{c2}(p)$, separating a superconducting
phase with fast spin fluctuations from a superconducting phase that
coexists with static magnetism. ZF-$\mu$SR measurements
indicate that superconductivity and static magnetism coexist on
a nanometer scale at $H \! = \! 0$, but only up to a carrier concentration
$p$ that is less than $p_{\rm MIC}$.
Above the dashed red curve, the static magnetism is uniformly
distributed throughout the sample.}
\label{fig22}
\end{figure*}

Kivelson {\it et al.} \cite{Kivelson:02} have extended the theory of Demler {\it et al.}
to account for the coupling between CuO$_2$ layers, and showed that in 
contrast to an isolated 2D vortex, a competing order parameter could be 
stabilized in an isolated 3D vortex line. This leads to a modification
of the phase diagram proposed by Demler {\it et al.}, such that the 
{\it true} QPT is to a phase in which a spatially inhomogeneous 
competing order coexists with superconductivity (see figure~\ref{fig21}b). 
The TF-$\mu$SR experiments on La$_{2-x}$Sr$_x$CuO$_4$ 
($0.12 \! < \! x \! < \! 0.166$) \cite{Kadono:04a,Sonier:06b}
and YBa$_2$Cu$_3$O$_y$ ($y \! < \! 6.57$) \cite{Sonier:06b,Miller:02}
strongly support the existence of this phase. The observed changes in 
the $\mu$SR line shape are consistent with the onset of {\it disordered}
static magnetism in and around weakly interacting vortices. 
These experiments imply that the competing order parameter 
stabilized at the QPT is the mean 
squared {\it local} magnetization. With increasing magnetic field,
stronger overlap of the magnetism around neighbouring vortices may lead to a 
co-operative bulk crossover to long-range magnetic order, as is 
apparently the case in La$_{1.856}$Sr$_{0.144}$CuO$_4$ \cite{Khaykovich:05}.
However, so far this has not been observed in YBa$_2$Cu$_3$O$_y$ where the
QPT identified by $\mu$SR is closer to the low-doping side of the 
superconducting `dome'. While the QPT in electron-doped cuprates
has not been accurately determined by $\mu$SR or neutron scattering, 
experiments to date suggest 
the coexistence phase is dominated by static magnetic order rather 
than disordered static magnetism---likely due to close proximity
of the AF and superconducting phases.

The TF-$\mu$SR experiments on La$_{2-x}$Sr$_x$CuO$_4$ and 
YBa$_2$Cu$_3$O$_y$ of Ref.~\cite{Sonier:06b} indicate that the QPT
in figure~\ref{fig21}b occurs at the critical doping below which
the temperature dependence of the normal-state
resistivity $\rho$ at $H \! > \! H_{c2}$ and low-$T$ 
changes from {\it metallic} behaviour ($d \rho/dT \! > \! 0$) to an 
unusual {\it insulating} behaviour ($d \rho/dT \! < \! 0$)
\cite{Boebinger:96,Fournier:98,Ono:00,Li:02,Dagan:04,Dagan:05}.
This low-temperature metal-to-insulator (MIC) crossover has
also been detected indirectly by thermal conductivity measurements 
as a function of magnetic field \cite{Sun:03,Hawthorn:03,Sun:04}.
Although the MIC is a generic property of high-$T_c$ cuprates,
it occurs at a non-universal carrier concentration.
The onset of static magnetism below the MIC detected by
TF-$\mu$SR is compatible with the localization of charge.
In zero-field, deviations from a linear dependence of the 
resistivity on $T$ are observed in underdoped samples,
including upturns of the resistivity with decreasing
temperature in very underdoped superconducting samples.
Theoretically, Marchetti {\it et al.} \cite{Marchetti:01,Marchetti:04}
have argued that the cause of the MIC
in both zero field and strong magnetic fields is the same,
and is due to the onset of short-range AF order \cite{Marchetti:01}.
This seems compatible with the ZF-$\mu$SR studies that have detected
static magnetism in underdoped samples.
          
A key prediction of the theory of Ref.~\cite{Kivelson:02} is that there 
is an `avoided' QCP at $H \! = \! 0$. In other words, the QCP lies at a 
lower doping than one expects from the extrapolated 
$H \! \rightarrow \! 0$ location of the QPT at finite magnetic
field. In La$_{2-x}$Sr$_x$CuO$_4$,
the spin freezing transition temperature $T_g$ at $H \! = \! 0$
deduced from ZF-$\mu$SR \cite{Niedermayer:98,Panagopoulos:02}  
and NQR \cite{Chou:93} experiments extrapolates to zero below
the critical doping for the MIC at $x \! \approx \! 0.16$
\cite{Boebinger:96,Sun:03,Hawthorn:03} (see figure~\ref{fig15}).
The same is true for YBa$_2$Cu$_3$O$_y$, where static magnetism
associated with Cu spins is not observed by ZF-$\mu$SR 
\cite{Kiefl:89,Sanna:04,Sonier:06b,Miller:02}
in samples immediately below the critical doping for the 
MIC at $y \! \approx \! 6.55$ \cite{Sun:04}.  

Figure~\ref{fig22} shows the generic magnetic phase diagram
suggested by ZF-$\mu$SR and TF-$\mu$SR experiments on cuprates. 
In this figure the QCP and `avoided' QCP are shown as solid and open circles, 
respectively. Note that the zero-field QCP here is distinct from the 
universal QCP near $p \! = \! 0.19$ that has been suggested by Tallon and 
others \cite{Panagopoulos:02,Tallon:01,Panagopoulos:05}.
In other words, there may in fact be two unrelated quantum critical points
under the superconducting `dome' that strongly influence the
physical properties of a large portion of the cuprate phase diagram.  
While neutron scattering and $\mu$SR measurements both support a 
generic phase diagram ruled by the close competition between
superconducting and magnetic orders, it is still to be determined whether 
{\it some} magnetism is also responsible for the `glue' that binds
the superconducting carriers.
        
\subsection{Acknowledgments}
The work presented here was supported by the Natural Science and Engineering Research
Council of Canada, and the Quantum Materials program of the 
Canadian Institute for Advanced Research.


\section*{References}

\begin{thebibliography}{21}
\bibitem{Sonier:00} Sonier J E, Brewer J H and Kiefl R F 2000 
{\it Rev. Mod. Phys.} {\bf 72} 769
\bibitem{Herlach:90} Herlach D {\it et al.} 1990 {\it Hyperfine Interactions} {\bf 63} 41
\bibitem{Sonier:97a} Sonier J E {\it et al.} 1997 Phys.~Rev.~Lett. 
{\bf 79} 1742
\bibitem{Golubov:94} Golubov A A and Hartmann U 1994 Phys.~Rev.~Lett. 
{\bf 72} 3602
\bibitem{Kirtley:99} Kirtley J R, Kogan V G, Clem J R and Moler K A 1999
Phys.~Rev.~B {\bf 59} 4343
\bibitem{Niedermayer:99} Niedermayer Ch {\it et al.} 1999 Phys.~Rev.~Lett.
{\bf 83} 3932
\bibitem{Morenzoni:04} Morenzoni E, Prokscha T, Suter A, Luetkens H and
Khasanov R 2004 {\it J. Phys.:Condens. Matter} {\bf 16} S4583
\bibitem{Salman:07} Salman Z {\it et al.} 2007 Phys.~Rev.~Lett. {\bf 98} 167001
\bibitem{Dawson:88} Dawson W K, Tibbs K, Weathersby S P, Boekema C and
Chan K-C B 1988 J.~Appl.~Phys. {\bf 64} 5809
\bibitem{Brewer:90} Brewer J H {\it et al.} 1990 {\it Hyperfine Interactions} {\bf 63} 177 
\bibitem{Schneider:93} Schneider J W {\it et al.} 1993 Phys.~Rev.~Lett. {\bf 71} 557
\bibitem{Chakhalian:97} Chakhalian J A {\it et al.} 1997 {\it Hyperfine Interactions}
{\bf 106} 245
\bibitem{Brandt:88b} Brandt E H 1988 Phys.~Rev.~B {\bf 37} 2349(R)
\bibitem{Brandt:91} Brandt E H 1991 Phys.~Rev.~Lett. {\bf 66} 3213
\bibitem{Yaouanc:98} Yaouanc A {\it et al.} 1998 {\it J. Phys.: Condens. Matter}
{\bf 10} 9791
\bibitem{Barford:88} Barford W and Gunn J M F 1988 {\it Physica} C {\bf 156} 515
\bibitem{Harshman:87} Harshman D R {\it et al.} 1987 Phys.~Rev.~B {\bf 36} 2386(R)
\bibitem{Uemura:88} Uemura Y J {\it et al.} 1988 Phys.~Rev.~B {\bf 38} 909(R)
\bibitem{Harshman:89} Harshman D R {\it et al.} 1989 Phys.~Rev.~B {\bf 39} 851(R)
\bibitem{Pumpin:90} P\"{u}mpin B {\it et al.} 1990 Phys.~Rev.~B {\bf 42} 8019
\bibitem{Uemura:89} Uemura Y J {\it et al.} 1989 Phys.~Rev.~Lett. {\bf 62} 2317
\bibitem{Uemura:91} Uemura Y J {\it et al.} 1991 Phys.~Rev.~Lett. {\bf 66} 2665
\bibitem{Yaouanc:97} Yaouanc A, Dalmas de R\'{e}otier P and Brandt E H 1997
Phys.~Rev.~B {\bf 55} 11107
\bibitem{Ando:02} Ando Y and Segawa K 2002 Phys.~Rev.~Lett. {\bf 88} 167005 
\bibitem{Wen:03} Wen H H {\it et al.} 2003 {\it Europhys. Lett.} {\bf 64} 790
\bibitem{Wang:03} Wang Y {\it et al.} 2003 {\it Science} {\bf 299} 86
\bibitem{Brandt:03} Brandt E H 2003 Phys.~Rev.~B {\bf 68} 054506
\bibitem{Celio:88} Celio M, Riseman T M, Kiefl R F, Brewer J H and
Kossler W J 1988 {\it Physica} C {\bf 153-155} 753
\bibitem{Pereg:04} Pereg-Barnea T {\it et al.} 2004 Phys.~Rev.~B {\bf 69} 184513
\bibitem{Zuev:05} Zuev Y, Kim M S and Lemberger T R 2005 Phys.~Rev.~Lett. {\bf 95} 137002
\bibitem{Broun:05} Broun D M {\it et al.} 2007 Phys.~Rev.~Lett. {\bf 99} 237003
\bibitem{Tallon:03} Tallon J L, Loram J W, Cooper J R, Panagopoulos C
and Bernhard C 2003 Phys.~Rev.~B {\bf 68} 180501(R)
\bibitem{Hardy:93} Hardy W N {\it et al.} 1993 Phys.~Rev.~Lett. {\bf 70} 3999
\bibitem{Sonier:94} Sonier J E {\it et al.} 1994 Phys.~Rev.~Lett. {\bf 72} 744
\bibitem{Luke:97} Luke G M {\it et al.} 1997 {\it Physica} C {\bf 282-287} 1465
\bibitem{Harlingen:95} Van Harlingen D J 1995 {\it Rev. Mod. Phys.} {\bf 67} 515
\bibitem{Amin:98} Amin M H S, Franz M and Affleck I 1998 Phys.~Rev.~B {\bf 58} 5848
\bibitem{Sonier:99b} Sonier J E {\it et al.} 1999 Phys.~Rev.~Lett. {\bf 83} 4156
\bibitem{Amin:00} Amin M H S, Franz M and Affleck I 2000 Phys.~Rev.~Lett. {\bf 84} 5864
\bibitem{Sonier:04c} Sonier J E 2004 {\it J. Phys.:Condens. Matter} {\bf 16} S4499
\bibitem{Callaghan:05} Callaghan F D, Laulajainen M, Kaiser C V and Sonier J E 2005
Phys.~Rev.~Lett. {\bf 95} 197001
\bibitem{Laulajainen:06} Laulajainen M, Callaghan F D, Kaiser C V and Sonier J E
2006 Phys.~Rev.~B {\bf 74} 054511
\bibitem{Laiho:06} Laiho R, Safonchik M and Traito
K B 2006 Phys.~Rev.~B {\bf 73} 024507
\bibitem{Laiho:07} Laiho R, Safonchik M and Traito
K B 2007 Phys.~Rev.~B {\bf 75} 174524
\bibitem{Landau:07} Landau I L and Keller H {\it arXiv:0704.1268v2}
\bibitem{Brandt:77} Brandt E H 1977 {\it J. Low Temp. Phys.} B {\bf 24} 709 
\bibitem{Brandt:88a} Brandt E H 1988 {\it J. Low Temp. Phys.} B {\bf 73} 355 
\bibitem{Brandt:92} Brandt E H 1992 {\it Physica} C {\bf 195} 1
\bibitem{Oliveira:98} de Oliveira I G and Thompson A M 1998
Phys.~Rev.~B {\bf 57} 7477
\bibitem{Laiho:05} Laiho R, L\"{a}hderanta E, Safonchik M and Traito
K B 2005 Phys.~Rev.~B {\bf 71} 024521
\bibitem{Riseman:95} Riseman T M {\it et al.} 1995 Phys.~Rev.~B {\bf 52}, 10569
\bibitem{Sonier:97c} Sonier J E {\it et al.} 1997 Phys.~Rev.~B {\bf 55} 11789
\bibitem{Kadono:01} Kadono R {\it et al.} 2001 Phys.~Rev.~B {\bf 63} 224520
\bibitem{Ohishi:02} Ohishi K {\it et al.} 2002 Phys.~Rev.~B {\bf 65} 140505
\bibitem{Price:02} Price A N {\it et al.} 2002 Phys.~Rev.~B {\bf 65} 214520
\bibitem{Clem:75} Clem J R 1975 {\it J. Low Temp. Phys.} {\bf 18} 427
\bibitem{Brandt:97} Brandt E H 1997 Phys.~Rev.~Lett. {\bf 78} 2208
\bibitem{Takanaka:71} Takanaka K 1971 {\it Prog. Theor. Phys.} {\bf 46} 1301 
\bibitem{Ichioka:96} Ichioka M, Hayashi N, Enomoto N and Machida K 1996 
Phys.~Rev.~B {\bf 53} 15316
\bibitem{Kogan:97a} Kogan V G {\it et al.} 1997 Phys.~Rev.~B {\bf 55} R8693
\bibitem{Kogan:97b} Kogan V G, Miranovi\'{c} P, Dobrosavljevi\'{c}-Gruji\'{c} Lj, Pickett W E
and Christen D K 1997 Phys.~Rev.~Lett. {\bf 79} 741
\bibitem{Nakai:02} Nakai N, Miranovi\'{c} P, Ichioka M and Machida K 2002 
Phys.~Rev.~Lett. {\bf 89} 237004
\bibitem{Luke:99} Luke G M {\it et al.} 2000 {\it Physica} B {\bf 289} 373
\bibitem{Sonier:04a} Sonier J E {\it et al.} 2004 Phys.~Rev.~Lett. {\bf 93} 017002
\bibitem{Kadono:06} Kadono K {\it et al.} 2006 Phys.~Rev.~B {\bf 74} 024513
\bibitem{Affleck:96} Affleck I, Franz M and Amin M H 1996 Phys.~Rev.~B {\bf 55} R704
\bibitem{Agterberg:98} Agterberg D F 1998 Phys.~Rev.~B {\bf 58} 14484
\bibitem{Heeb:99} Heeb R and Agterberg D F 1999 Phys.~Rev.~B {\bf 59} 7076
\bibitem{Sosolik:03} Sosolik C E {\it et al.} 2003 Phys.~Rev.~B {\bf 68} 140503(R)
\bibitem{Giamarchi:95} Giamarchi T and Le Doussal P 1995 Phys.~Rev.~B {\bf 52} 1242
\bibitem{Dalmas:04} Dalmas de R\'{e}otier P, Gubbens P C M and
Yaouanc A 2004 {\it J. Phys.:Condens. Matter} {\bf 16} S4687
\bibitem{Divakar:04} Divakar U {\it et al.} 2004 Phys.~Rev.~Lett. {\bf 92} 237004
\bibitem{Menon:06} Menon G I {\it et al.} 2006 Phys.~Rev.~Lett. {\bf 97} 177004
\bibitem{Koshelev:96} Koshelev A E, Glazman L I and Larkin A I 1996 Phys.~Rev.~B {\bf 53}, 2786
\bibitem{Menon:99} Menon G I {\it et al.} 1999 Phys.~Rev.~B {\bf 60} 7207 
\bibitem{Harshman:93} Harshman D R {\it et al.} 1993 Phys.~Rev.~B {\bf 47} 2905
\bibitem{Harshman:04} Harshman D R {\it et al.} 2004 Phys.~Rev.~B {\bf 69} 174505
\bibitem{Deak:95} Deak J, Hou L, Metcalf P and McElfresh M 1995 {it Phys. Rev.} B {\bf 51} 705(R)
\bibitem{Sefrioui:99} Sefrioui Z {\it et al.} 1999 Phys.~Rev.~B {\bf 60} 15423
\bibitem{Dulic:01} Duli\'{c} D {\it et al.} 2001 Phys.~Rev.~Lett. {\bf 86}, 4660
\bibitem{Ichioka:99a} Ichioka M, Hasegawa A and Machida K 1999 Phys.~Rev.~B {\bf 59} 184
\bibitem{Ichioka:99b} Ichioka M, Hasegawa A and Machida K 1999 Phys.~Rev.~B {\bf 59} 8902
\bibitem{Caroli:64} Caroli C, de Gennes P G and Matricon J 1964 {\it Phys. Lett.} {\bf 9} 307
\bibitem{Pottinger:93} P\"{o}ttinger B and Klein U 1993 Phys.~Rev.~Lett. {\bf 70} 2806
\bibitem{Tesanovic:94} Dukan S and Tesanovi\'{c} Z 1994 Phys.~Rev.~B {\bf 49} 13017
\bibitem{Tesanovic:98} Tesanovi\'{c} Z and Sacramento P 1998 Phys.~Rev.~Lett. {\bf 80} 1521
8902
\bibitem{Kogan:04} Kogan V G and Zhelezina N V 2004 Phys.~Rev.~B {\bf 71} 134505
\bibitem{DeBeer:06} DeBeer-Schmitt L, Dewhurst C D, Hoogenboom B W, Petrovic C
and Eskildsen M R 2006 Phys.~Rev.~Lett. {\bf 97} 127001
\bibitem{Nakai:06} Nakai N, Miranovi\'{c} P, Ichioka M and Machida K 2006 Phys.~Rev.~B {\bf 73} 172501
\bibitem{Fetter:69} Parks R D (ed) 1969 in {\it Superconductivity} vol~2
(New York: Marcel Dekker Inc.) p~817
\bibitem{Volovik:93} Volovik G E 1993 {\it JETP Lett.} {\bf 58} 469
\bibitem{Sonier:99a} Sonier J E, Hundley M F, Thompson J D and Brill J W 1999 
Phys.~Rev.~Lett. {\bf 82} 4914
\bibitem{Nakai:04} Nakai N, Miranovi\'{c} P, Ichioka M and Machida K 2004 Phys.~Rev.~B {\bf 70} 100503 
\bibitem{Sonier:06a} Sonier J E, Hundley M F and Thompson J D 2006 
Phys.~Rev.~B {\bf 73} 132504
\bibitem{Boaknin:01} Boaknin E {\it et al.} 2001 Phys.~Rev.~Lett. {\bf 87} 237001
\bibitem{Chiao:99} Chiao M, Hill R W, Lupien C, Popic B, Gagnon R and Taillefer L 1999 
Phys.~Rev.~Lett. {\bf 82} 2943
\bibitem{Kubert:98} K\"{u}bert C and Hirschfeld P J 1998 Phys.~Rev.~Lett. {\bf 80} 4963
\bibitem{Vekhter:99} Vekhter I and Houghton A 1999 Phys.~Rev.~Lett. {\bf 83} 4626
\bibitem{Boaknin:03} Boaknin E {\it et al.} 2003 Phys.~Rev.~Lett. {\bf 90} 117003
\bibitem{Lowell:70} Lowell J and Sousa J B 1970 {\it J. Low Temp. Phys.} {\bf 3} 65
\bibitem{Yokoya:01} Yokoya T {\it et al.} 2001 {\it Science} {\bf 294} 2518
\bibitem{Rodrigo:04} Rodrigo J G and Vieira S 2004 {\it Physica} C {\bf 404} 306 
\bibitem{Carrington:06} Fletcher J D {\it et al.} 2007 Phys.~Rev.~Lett. {\bf 75} 045019
\bibitem{Nakai:02b} Nakai N, Ichioka M and Machida K 2002 J.~Phys.~Soc.~Jpn.
{\bf 71} 23
\bibitem{Koshelev:03} Koshelev A E and Golubov A A 2003 Phys.~Rev.~Lett. {\bf 90} 177002
\bibitem{Dahm:04} Dahm T, Graser S and Schopohl N 2004 {\it Physica} C {\bf 408-410} 336
\bibitem{Ichioka:04} Ichioka M, Machida K, Nakai N and Miranovi\'{c} P 2004
Phys.~Rev.~B {\bf 70} 144508
\bibitem{Eskildsen:02} Eskildsen M R {\it et al.} 2002 Phys.~Rev.~Lett. {\bf 89} 187003
\bibitem{Klein:06} Klein T, Lyard L, Marcus J, Holanova Z and Marcenat C 2006
Phys.~Rev.~B {\bf 73} 184513
\bibitem{Sologubenko:02} Sologubenko A V, Jun J, Kazakov S M, Karpinski J and Ott H R
2002 Phys.~Rev.~B {\bf 66} 014504
\bibitem{Seyfarth:05} Seyfarth G {\it et al.} 2005 Phys.~Rev.~Lett. {\bf 95} 107004
\bibitem{Seyfarth:06} Seyfarth G {\it et al.} 2006 Phys.~Rev.~Lett. {\bf 97} 236403
\bibitem{Atkinson:07} Atkinson W A, private communication.
\bibitem{Brown:04} Brown S P {\it et al.} 2004 Phys.~Rev.~Lett. {\bf 92} 067004
\bibitem{Hill:04} Hill R W {\it et al.} 2004 Phys.~Rev.~Lett. {\bf 92} 027001
\bibitem{Shulga:98} Shulga S V {\it et al.} 1998 Phys.~Rev.~Lett. {\bf 80} 1730
\bibitem{Doh:99} Doh H, Sigrist M, Cho B K and Lee S -I 1999 
Phys.~Rev.~Lett. {\bf 83} 5350
\bibitem{Mukhop:05} Mukhopadhyay S, Sheet G, Raychaudhuri P and Takeya H
2005 Phys.~Rev.~B {\bf 72} 014545
\bibitem{Huang:06} Huang C L {\it et al.} 2006 Phys.~Rev.~B {\bf 73} 012502
\bibitem{Nakagawa:98} Nakagawa H, Takamasu T, Miura N and Enomoto Y 1998 
{\it Physica} B {\bf 246-247} 429
\bibitem{Kadono:04c} Kadono R 2004 {\it J. Phys.:Condens. Matter} {\bf 16} S4421
\bibitem{Kiefl:89} Kiefl R F {\it et al.} 1989 Phys.~Rev.~Lett. {\bf 63} 2136
\bibitem{Weidinger:89} Weidinger A {\it et al.} 1989 Phys.~Rev.~Lett. {\bf 62} 102
\bibitem{Heffner:89} Heffner R H and Cox D L 1989 Phys.~Rev.~Lett. {\bf 63} 2538
\bibitem{Niedermayer:98} Niedermayer Ch {\it et al.} 1998 Phys.~Rev.~Lett. {\bf 80} 3843
\bibitem{Panagopoulos:02} Panagopoulos C {\it et al.} 2002 Phys.~Rev.~B {\bf 66} 064501
\bibitem{Sonier:01} Sonier J E {\it et al.} 2001 {\it Science} {\bf 292} 1692
\bibitem{Kanigel:02} Kanigel A {\it et al.} 2002 Phys.~Rev.~Lett. {\bf 88} 137003
\bibitem{Sanna:04} Sanna S, Allodi G, Concas G, Hillier A D and De Renzi R 2004 
Phys.~Rev.~Lett. {\bf 93} 207001
\bibitem{Miller:06} Miller R I {\it et al.} 2006 Phys.~Rev.~B {\bf 73} 144509
\bibitem{Basov:95} Basov D N {\it et al.} 1995 Phys.~Rev.~Lett. {\bf 74} 598
\bibitem{Lee:05} Lee Y-S, Segawa K, Ando Y amd Basov D N 2005 Phys.~Rev.~Lett. {\bf 94} 137004
\bibitem{Atkinson:95} Atkinson W A and Carbotte J P 1995 Phys.~Rev.~B {\bf 52} 10601
\bibitem{Xiang:96} Xiang T and Wheatley J M 1996 Phys.~Rev.~Lett. {\bf 76} 134
\bibitem{Grevin:00} Gr\'{e}vin B, Berthier Y and Collin G 2000 Phys.~Rev.~Lett. 
{\bf 85} 1310
\bibitem{Sonier:02} Sonier J E {\it et al.} 2002 Phys.~Rev.~B {\bf 66} 134501
\bibitem{Cho:92} Cho J H, Borsa F, Johnston D C and Torgeson D R 1992 
Phys.~Rev.~B {\bf 46} 3179
\bibitem{Chou:93} Chou F C {\it et al.} 1993 Phys.~Rev.~Lett. {\bf 71} 2323
\bibitem{Julien:99} Julien M -H {\it et al.} 1999 Phys.~Rev.~Lett. {\bf 83} 604
\bibitem{Birgeneau:06} Birgeneau R J, Stock C, Tranquada J M and Yamada K 2006
J.~Phys.~Soc.~Jpn. {\bf 75} 111003
\bibitem{Wakimoto:99} Wakimoto S {\it et al.} 1999 Phys.~Rev.~B {\bf 60} R769
\bibitem{Matsuda:00} Matsuda M {\it et al.} 2000 Phys.~Rev.~B {\bf 62} 9148
\bibitem{Wakimoto:01} Wakimoto S {\it et al.} 2001 Phys.~Rev.~B {\bf 63} 172501
\bibitem{Shraiman:89} Shraiman B I and Siggia E D 1989 Phys.~Rev.~Lett.
{\bf 62} 1564
\bibitem{Luscher:07} L\"{u}scher A, Milstein A I and Sushkov 2007
Phys.~Rev.~Lett. {\bf 98} 037001 
\bibitem{Sanna:05} Sanna S, Allodi G, Concas G and De Renzi R 2005 
{\it J. Supercond.} {\bf 18} 769
\bibitem{Stock:06} Stock C {\it et al.} 2006 Phys.~Rev.~B {\bf 73} 100504(R)
\bibitem{Curro:00} Curro N J, Milling C, Haase J and Slichter C P 2000 Phys.~Rev.~B {\bf 62} 3473 
\bibitem{Mitrovic:01} Mitrovi\'{c} V F {\it et al.} 2001 Nature {\bf 413} 501
\bibitem{Mitrovic:03} Mitrovi\'{c} V F {\it et al.} 2003 Phys.~Rev.~B {\bf 67} 220503(R)
\bibitem{Kakuyanagi:02} Kakuyanagi K, Kumagai K and Matsuda Y 2002 Phys.~Rev.~B {\bf 65} 060503  
\bibitem{Kakuyanagi:03} Kakuyanagi K, Kumagai K, Matsuda Y and Hasegawa M 2003 
Phys.~Rev.~Lett. {\bf 90} 197003  
\bibitem{Lake:01} Lake B {\it et al.} 2001 {\it Science} {\bf 291} 1759
\bibitem{Kadono:04a} Kadono R {\it et al.} 2004 Phys.~Rev.~B {\bf 69} 104523
\bibitem{Lake:02} Lake B {\it et al.} 2002 {\it Nature} {\bf 415} 299
\bibitem{Katano:00} Katano S, Sato M, Yamada K, Suzuki T and Fukase T 2000
Phys.~Rev.~B {\bf 62} 14677(R)
\bibitem{Khaykovich:05} Khaykovich B {\it et al.} 2005 Phys.~Rev.~B {\bf 71} 220508(R)
\bibitem{Sonier:06b} Sonier J E {\it et al.} 2007 Phys.~Rev.~B {\bf 76} 064522
\bibitem{Lee:93} Lee SL {\it et al.} 1993 Phys.~Rev.~Lett. {\bf 71} 3862
\bibitem{Lake:05} Lake B {\it et al.} 2005 {\it Nature Materials} {\bf 4} 658
\bibitem{Savici:05} Savici A T {\it et al.} 2005 Phys.~Rev.~Lett. {\bf 95} 157001
\bibitem{Machtoub:05} Machtoub L H, Keimer B and Yamada K 2005 
Phys.~Rev.~Lett. {\bf 94} 107009
\bibitem{Miller:02} Miller R I {\it et al.} 2002 Phys.~Rev.~Lett. {\bf 88} 137002
\bibitem{Sonier:03} Sonier J E {\it et al.} 2003 Phys.~Rev.~Lett. {\bf 91} 147002
\bibitem{Sonier:04b} Sonier J E {\it et al.} 2004 {\it Physica} C {\bf 408} 783
\bibitem{Kadono:04b} Kadono R {\it et al.} 2004 J.~Phys.~Soc.~Jpn. {\bf 73} 2944
\bibitem{Kadono:05} Kadono R {\it et al.} 2005 J.~Phys.~Soc.~Jpn. {\bf 74} 2806
\bibitem{Fujita:04} Fujita M, Matsuda M, Katano S and Yamada K 2004 
Phys.~Rev.~Lett. {\bf 93} 147003
\bibitem{Kadono:03} Kadono R {\it et al.} 2003 J.~Phys.~Soc.~Jpn. {\bf 72} 2955
\bibitem{Kang:05} Kang H J {\it et al.} 2005 Phys.~Rev.~B {\bf 71} 214512
\bibitem{Demler:01} Demler E, Sachdev S and Zhang Y 2001 Phys.~Rev.~Lett. {\bf 87} 067202
\bibitem{Park:06} Park T {\it et al.} 2006 {\it Nature} {\bf 440} 65
\bibitem{Kivelson:02} Kivelson S A, Lee D -H, Fradkin E and Oganesyan V 2002
Phys.~Rev.~B {\bf 66} 144516
\bibitem{Khaykovich:02} Khaykovich B {\it et al.} 2002 Phys.~Rev.~B {\bf 66} 014528
\bibitem{Boebinger:96} Boebinger G S {\it et al.} 1996 Phys.~Rev.~Lett. {\bf 77} 5417  
\bibitem{Fournier:98} Fournier P {\it et al.} 1998 Phys.~Rev.~Lett. {\bf 81} 4720
\bibitem{Ono:00} Ono S {\it et al.} 2000 Phys.~Rev.~Lett. {\bf 85} 638
\bibitem{Li:02} Li S Y {\it et al.} 2002 Phys.~Rev.~B {\bf 65} 224515
\bibitem{Dagan:04} Dagan Y {\it et al.} 2004 Phys.~Rev.~Lett. {\bf 92} 167001  
\bibitem{Dagan:05} Dagan Y {\it et al.} 2005 Phys.~Rev.~Lett. {\bf 94} 057005    
\bibitem{Sun:03} Sun X F, Komiya S, Takeya J and Ando Y 2003 
Phys.~Rev.~Lett.{\bf 90} 117004
\bibitem{Hawthorn:03} Hawthorn D G {\it et al.} 2003 Phys.~Rev.~Lett. {\bf 90} 197004
\bibitem{Sun:04} Sun X F, Segawa K and Ando Y 2004 Phys.~Rev.~Lett. {\bf 93} 107001
\bibitem{Marchetti:01} Marchetti P A, Su Z B and Yu L 2001 Phys.~Rev.~Lett. {\bf 86} 3831  
\bibitem{Marchetti:04} Marchetti P A, De Leo L, Orso G, Su Z B and Yu L 2004 
Phys.~Rev.~B {\bf 69} 024527
\bibitem{Tallon:01} Tallon J L and Loram J W 2001 {\it Physica} C {\bf 349} 53  
\bibitem{Panagopoulos:04} Panagopoulos C {\it et al.} 2004 Phys.~Rev.~B {\bf 69} 144510
\bibitem{Panagopoulos:05} Panagopoulos C and Dobrosavljevi\'{c} V 2005 Phys.~Rev.~B {\bf 72} 014536
\end{thebibliography}
\end{document}